\documentclass[LectureNotes]{SciPost}
\usepackage{epsfig,amsfonts,cite}
\usepackage[greek,english]{babel}
\def\coppa{{\fontencoding{LGR}\fontfamily{cmr}\selectfont\textqoppa}}

\usepackage{pstricks}
\usepackage{amsmath}
\usepackage{empheq}
\usepackage{pstricks}
\usepackage{amssymb}
\usepackage{mathrsfs}
\usepackage{mathtools}
\usepackage{cancel}
\usepackage{accents}
\usepackage{graphicx}
\usepackage{dsfont}
\usepackage{lmodern}
\usepackage{orcidlink}
\usepackage[normalem]{ulem}

\DeclareRobustCommand{\augiefamily}{%
  \fontfamily{augie}\fontseries{m}\fontshape{n}\selectfont}
\DeclareTextFontCommand{\textaugie}{\augiefamily}

\def\be{\begin{equation}}
\def\ee{\end{equation}}
\def\bea{\begin{eqnarray}}
\def\eea{\end{eqnarray}}
\def\bpar{\left(\!\!\begin{array}}
\def\epar{\end{array}\!\!\right)}
\def\bdar{\left|\!\!\begin{array}}
\def\edar{\end{array}\!\!\right|}
\def\barr{\begin{array}}
\def\earr{\end{array}}
\def\btab{\begin{tabular}}
\def\etab{\end{tabular}}
\def\nnb{\nonumber}
\def\vac#1{{\bf #1}}
\def\bnabla{\pmb\nabla}
\def\<{\langle}
\def\>{\rangle}

\def\half{{\textstyle\frac 12}}
\def\quarter{{\textstyle\frac 14}}
\def\MFT{{\scriptscriptstyle\rm MFT}}
\def\G{{\scriptscriptstyle\rm G}}

\definecolor{antiquefuchsia}{rgb}{0.57, 0.36, 0.51}
\definecolor{burgundy}{rgb}{0.5, 0.0, 0.13}

\definecolor{MyDarkGreen}{rgb}{0.02,0.60,0.06}
\definecolor{MyDarkGreen}{rgb}{1.00,0.00,0.00}

\makeatletter
\newcommand{\sbullet}{%
  \hbox{\fontfamily{lmr}\fontsize{.4\dimexpr(\f@size pt)}{0}\selectfont\textbullet}}

\makeatother

\definecolor{burgundy}{rgb}{0.5, 0.0, 0.13}

\def\mathcoppa{\hbox{\foreignlanguage{greek}{\coppa}}}

\def\qq{{\hbox{\foreignlanguage{greek}{\coppa}}}}
\def\qqq{{\hbox{\foreignlanguage{greek}{\footnotesize\coppa}}}}

\usepackage{upgreek}

\def\Rvsd#1{\textcolor{black}{#1}}

\begin{document}

\begin{center}{\Large \textbf{
Phase transitions above the upper critical dimension
}}\end{center}

\begin{center}
Bertrand Berche\orcidlink{0000-0002-4254-807X}\textsuperscript{a,d}, Tim Ellis\orcidlink{0000-0003-4713-7875}\textsuperscript{b,d}, Yurij Holovatch\orcidlink{0000-0002-1125-2532}\textsuperscript{c,d,b,e},  Ralph Kenna\orcidlink{0000-0001-9990-4277}\textsuperscript{b,d}
\end{center}

\begin{center}
{\bf a} Laboratoire de Physique et Chimie Th\'eoriques,\\ 
Universit\'e de Lorraine - CNRS, UMR 7019, Nancy,
B.P. 70239, F-54506 Vand\oe uvre les Nancy, France
\\
{\bf b} Centre for Fluid and Complex Systems,\\ 
Coventry University, Coventry, CV1 5FB, United Kingdom \\
{\bf c} Institute for Condensed Matter Physics, \\ National Acad. Sci. of Ukraine,  UA–79011 Lviv, Ukraine \\
{\bf d} ${\mathbb L}^4$ Collaboration \& Doctoral College for the Statistical Physics of Complex Systems, \\ Leipzig-Lorraine-Lviv-Coventry, Europe \\
{\bf e}  Complexity Science Hub Vienna, 1090 Vienna, Austria
\end{center}

\begin{center}
\today
\end{center}


\section*{Abstract}
{\bf 
{These lecture notes provide an overview of the renormalization group (RG) as a successful framework to understand critical phenomena above the upper critical dimension $d_{\rm uc}$.
After an introduction to the scaling picture of continuous phase transitions, we discuss the apparent failure of the Gaussian fixed point to capture scaling for Landau mean-field theory, which should hold in the thermodynamic limit above $d_{\rm uc}$. 
We recount how Fisher's dangerous-irrelevant-variable formalism applied to  thermodynamic functions partially repairs the situation but at the expense of hyperscaling and finite-size scaling, both of which were, until recently, believed not to apply above $d_{\rm uc}$.
We recall limitations of various attempts to match the RG with analytical and numerical results for Ising systems.
We explain how the extension of dangerous irrelevancy to the correlation sector is key to marrying the above concepts into a comprehensive RG scaling picture that renders hyperscaling and finite-size scaling valid in all dimensions.  
We collect what we believe is the current status of the theory, including some new insights and results. 
This paper is in grateful memory of Michael Fisher who introduced many of the concepts discussed and who, half a century later, contributed to their advancement. }
}

\vspace{10pt}
\noindent\rule{\textwidth}{1pt}
\tableofcontents\thispagestyle{fancy}
\noindent\rule{\textwidth}{1pt}
\vspace{10pt}

\section{Prelude: Definitions and notations}
\label{SecNotations}

Without loss of generality, we present some of this exposition  within the framework of the Ising model as it is the original and most fundamental spin model {and the one generically used in textbooks}. 
Generalization to other models is straightforward and implicit.
{Notwithstanding this,}  we find it illustrative to present most of the work in the context of $\phi^n$ theory. 
For clarity, we first specify the notations used and some specific vocabulary. 
This section can be skipped  by readers well versed in statistical physics theories of critical phenomena.

The {Ising} model is defined on a hypercubic lattice $\Lambda$, of volume $\ell^d$, and edge length $\ell=L a$ in space of dimension $d$. 
Here $L$ is a number of unit cells in any direction of the lattice and $a$ is their {individual} length. 
The sites of $\Lambda$ are assigned coordinates $\vac x_i=(x_{i_1},x_{i_2},\dots,x_{i_d})$ with $x_{i_n}=1,2,\dots,L$. At each lattice site sits a dimensionless Ising spin variable $s_{\vac x_i}=\pm 1$. 
In the case of short range interactions, these local degrees of freedom interact between nearest neighbours, $(\vac x_i,\vac x_j)$ through the so-called ``exchange interaction'', as well as with an external magnetic field, so that the Hamiltonian is
\be
{\mathscr H}=-\sum_{(\vac x_i,\vac x_j)}Js_{\vac x_i}s_{\vac x_j}-\sum_{\vac x_i}Hs_{\vac x_i}.
\ee
Here, the interaction coupling $J$ and the magnetic field $H$ have  dimensions of energy. 
They are considered uniform for the simplest version of the model {presented} here.
Generalisation to long-range models is straightforward.

{This} Ising model above $d=1$ exhibits a phase transition between an ordered state at low temperature and a disordered state at high temperature. 
At the transition, which occurs at finite temperature, the physical properties become singular and this is what we are interested in. 
An equilibrium order parameter discriminates between the two phases. 
It is zero above the transition in the thermodynamic limit and non-zero below.
{Thermodynamic properties such as these} are deduced from the partition function 
\be
Z(\beta,H)=\sum_{\{s_{\vac x_i}\}} e^{-\beta{\mathscr H}}
\label{ZZZ}
\ee
where the sum is over all the spin configurations and $\beta$ is the inverse temperature.
The free energy  is
\be
F_\infty(\tau,h)=-\beta^{-1}\ln Z_\infty(\beta,H)
\ee 
where $\tau=(\beta J-\beta_cJ)/(\beta_c J)$ and $h=\beta H$ with $\beta_c$ the inverse critical temperature.
The subscript $\infty$ here, and throughout, indicates that the observable is considered in the thermodynamic limit. 
Various thermodynamic quantities are obtained through  derivatives of the free energy density $f_\infty=V^{-1}F_\infty$ wrt the temperature and/or the magnetic field. {(Here $V=L^d$.)} E.g. the magnetization,  internal energy,  susceptibility and  specific heat, are
\bea
&&m_\infty(\tau,h)=\frac{\partial f_\infty(\tau,h)}{\partial h},\\
&&e_\infty(\tau,h)=\frac{\partial f_\infty(\tau,h)}{\partial \tau},\\
&&\chi_\infty(\tau,h)=\frac{\partial^2 f_\infty(\tau,h)}{\partial h^2},\\
&&c_\infty(\tau,h)=\frac{\partial^2 f_\infty(\tau,h)}{\partial \tau^2}.
\eea
 In these definitions, dimensional factors (and signs) are not considered as they are not needed for our purpose.

In practical numerical simulations, the equilibrium properties of the model are defined after averaging with  {appropriate} weights. E.g. for a physical quantity $Q$, defined in terms of the local degrees of freedom,
\be
\langle Q \rangle=\frac 1Z\sum_{\{s_{\vac x_i}\}}Q(\{s_{\vac x_i}\}) e^{-\beta{\mathscr H}}.
\ee
Numerically, the sum over  spin configurations  is sampled via a random walk in  configuration space and the weighted average is produced by specific transition rules among the configurations sampled. 
These transition rules and the detail of the building of the configurations are properly defined for each type of algorithm (Metropolis, cluster algorithms, reweighting algorithms, Fukui-Todo algorithms, etc.). This results usually in a list of values for the average ``spin'' and average energy, \be
S=\sum_{\vac x_i} s_{\vac x_i},\qquad E=\sum_{(\vac x_i,\vac x_j)}s_{\vac x_i}s_{\vac x_j},
\ee
{so that}
for each configuration {$S$ and $E$ take values $S_n$, $E_n$ viz}
\bea
n & S_n& E_n\nnb\\
1&S_1& E_1\nnb\\
2&S_2& E_2\nnb\\
3&S_3& E_3\nnb\\
&\vdots&\vdots \, .
\eea
The equilibrium value of an observable $Q(S,E)$ is then calculated by  averaging the values in the list,
\be
\langle Q\rangle = \frac{1}{N_{\rm conf}}\sum_{n=1}^{N_{\rm conf}}Q(S_n,E_n).
\ee
Here, ${N_{\rm conf}}$ is the number of useful {Monte Carlo (MC)} iterations after thermalization and after other possible  transient regimes have been discarded. 
In order to simplify later definitions, we  denote the average over the MC iterations with brackets:
\be
\langle (\dots)\rangle= \frac{1}{N_{\rm conf}}\sum_{n=1}^{N_{\rm conf}} (\dots)\, . \ee

The equilibrium spontaneous magnetization (in zero magnetic field) requires special precaution.
The values of $S_n$ have the same probability to be positive or negative. 
This is because of the invariance of the model under Z$_2$ symmetry which sends $S_{\vac x_i}\to -S_{\vac x_i} \forall i$. 
The non-zero value of the equilibrium spontaneous magnetization in a real material results from spontaneous symmetry breaking in the thermodynamic limit $L\to\infty$. In MC simulations of finite systems, the symmetry must be artificially broken and this is usually done by using absolute values. 
The   equilibrium spontaneous magnetization (density) for a system of size $L$ is therefore defined as
 \be
m_L 
= L^{-d} \langle |S_L|\rangle.
\ee
The  susceptibility is defined by the corresponding second moment,
\be
\chi_L =  L^{-d}(\langle S_L^2 \rangle- \langle |S_L|\rangle^2).
\ee
The subtracted term here is needed only in the ordered phase to get rid of the non-zero average magnetization there and delivers a vanishing susceptibility in the limit $T\to 0$.
The energy and specific heat are defined accordingly:
 \bea
&& e_L 
= L^{-d} \langle E_L\rangle\, ,\\
&&c_L =  L^{-d}(\langle E_L^2 \rangle- \langle E_L\rangle^2).
\eea
We use the term ``magnetic sector" to refer to quantities defined solely in terms of the $S_n$'s and we use the term ``energy sector" for  those defined solely in terms of the $E_n$'s. 
Quantities in the energy sector are often more difficult to analyze due to the presence of strong non-singular contributions at the transition.

In establishing the list over MC iterations, we may need more information (for example the individual values of the local spins for each configuration) if local quantities are to be calculated. This is the case if we want to measure the order parameter profile:
\be
m(\vac x_i)=\langle| s_{\vac x_i} |\rangle.\label{17}
\ee
 The (connected) correlation function  and the correlation length are defined from this local order parameter,
\be
g_L(\vac x, \tau,h)=\<m(0,t,h)m(\vac x,\tau,h)\>-\<m(0,\tau,h)\> \<m(\vac x,\tau,h)\>
\sim e^{-|\vac x|/\xi_L(\tau,h)}.\label{eq-xicorp}
\ee
\Rvsd{In this expression,  large $|\vac x|$ is understood, since to define $\xi$, one takes the limit $\log g(|\vac x|)/\log|\vac x|$ for $|\vac x|\to\infty$. Moreover, a power prefactor $|\vac x|^p$ is expected in the rhs of (\ref{eq-xicorp}).}

It is useful to define local quantities 
in terms of 
Fourier modes $\psi_{\vac k}$.
\Rvsd{In a system with periodic boundary conditions (PBC) the Fourier modes are  plane waves
\be
\psi_{\vac k}=L^{-d/2}e^{i\vac k\cdot\vac x}
\ee
and the boundary conditions select the  wave vectors  $\vac k=(2\pi/L)\vac n$, $\vac n\in\mathbb Z^d$. In a system with free boundary conditions (FBC), 
the Fourier modes are sine waves
\be
\psi_{\vac k}=\sqrt{2/L}\prod_{\alpha=1}^d\sin k_\alpha x_\alpha
\ee
with wave vectors
$k_\alpha=n_\alpha \pi/(L+1)$, $n_\alpha=1,2,\dots,L$.
Local quantities, like $s_{\vac x_i}$, now read as}
\be
s_{\vac x_i}=\sum_{\vac k\in{\cal K}}\tilde s_{\vac k}\psi_{\vac k}.\label{11}
\ee
The ``single-mode equilibrium magnetization'' and the corresponding susceptibility are  thus
\bea
&&m_{\vac k}=\langle |\tilde s_{\vac k}|\rangle,\label{e19}\\
&&\chi_{\vac k}=\langle |\tilde s_{\vac k}|^2\rangle-\langle |\tilde s_{\vac k}|\rangle^2.
\eea

We {denote by} ${\cal K}$ the set of all $\vac k$-modes. 
If $m(\vac x)$ possesses certain symmetries we can {partition} the modes in two sets, ${\cal K}={\cal Q}\cup{\cal G}$ where the modes pertaining to ${\cal Q}$ possess the same symmetry as $m(\vac x)$ and ${\cal G}$ is the orthogonal set. 
Then, we refer to the modes which mostly contribute to  Eq.~(\ref{17}), $\vac k\in{\cal Q}$, as $Q$-modes. 
The modes in ${\cal G}$ are called $G$-modes.
{The reason for this notation will become clear later on. Suffice to say for now that $G$ refers to ``Gaussian'' and $Q$ refers to a seemingly inert parameter that turns out to play a crucial role in understanding critical phenomena in high dimensions.}

 Translation invariance \Rvsd{of a periodic system} requires a uniform equilibrium  profile $m(\vac x)={\rm constant}$, and only the zero mode for which $\vac n=0$ has the symmetry of the uniform profile. It follows that 
\be
m(\vac x)=m_0=\langle| \tilde s_{\vac k=0} |\rangle
\ee
and ${\cal Q}$ contains only the zero mode while ${\cal G}$  contains all the other modes. A typical element of ${\cal G}$ is for example $\vac k_G=(2\pi/L) (1,0,0,\dots,0)$. Although $\vac k_G$ does not contribute to the equilibrium magnetization, it is interesting in some circumstances to analyze the behaviour of quantities like 
$m_{\vac k_G}$.

\Rvsd{In a system with free boundaries,} the equilibrium profile
$m(\vac x)$ is an even function wrt the center of the system, and only the modes for which the $n_\alpha$'s are all odd share this symmetry. These are the $Q$-modes, a typical representative of which is 
$\vac k_Q=(\pi/(L+1)) (1,1,1,\dots,1)$. All the other modes are $G$-modes and a typical $G$-mode is 
$\vac k_G=(\pi/(L+1)) (2,1,1,\dots,1)$. There is no zero mode for FBC's.

Having set up this framework, and established the notation, we next proceed to introduce the problem at hand ---  scaling for critical phenomena in high dimensions.

\section{Introduction: Second-order phase transitions and the scaling picture}
\label{sec1}

{In this paper}, we want to revisit an old question, namely {scaling in high dimensions and} the role of dangerous irrelevant variables (DIV) {---} a very smart stratagem introduced by Michael Fisher in the 1980's.
Fisher introduced the concept to reconcile the celebrated renormalization group (RG) picture of critical phase transitions with  results from  mean-field theory (MFT).
The latter holds above the upper critical dimension where fluctuations are known to become irrelevant.
We are not aware of any journal article by Fisher on this problem, except a short exposition in Ref.\cite{Privman1983}, but we can find the argument developed in the wonderful appendix D of a course that he gave in Stellenbosch in 1982\cite{FisherStellenbosch}. There he  concluded by saying
\begin{quotation}
{\em 
The moral of this story is that the standard scaling relations for critical exponents depend, in their derivation, on assumptions, usually left tacit, about the  nonsingular or nonvanishing behaviour of various scaling functions and their arguments. In many cases  these assumptions are valid and may be confirmed by explicit calculation (or other knowledge) but in certain circumstances they may fail, in which case an exponent relation may change its form. Other nontrivial cases of dangerous irrelevant variables are known so that phenomenon, although not common, is not truly exceptional.}
\end{quotation}
Earlier, in the same appendix, he  made the important   statement:
\begin{quotation}
{\em 
Notice [\dots] that the renormalization group framework has been preserved intact: the only flaw in the original argument was a failure to recognize and allow for possible singular behaviour of the scaling function.}
\end{quotation}

{We revisit these  statements at the end of this paper and adapt them for our main conclusions which are not dissimilar. 
In a nutshell, what Fisher and subsequent authors did for free energy and its derivatives, we do for the correlation function and correlation length. 
{This is needed for the RG framework to be ``preserved intact'' for finite-size systems too. }}

{We} consider various systems [e.g., $O(N)$ models, percolation, and tricritical points] with PBC's, for which we believe that the theory is fully settled.
After that, we address the case of FBC's  (Section \ref{sec9}). 
There, the results as we will see  are still incomplete. 

These notes should not be considered as a review. 
We do not purport to give a full account of the vast literature published over  almost half a century, but we will refer to those papers that we consider as being the most {important} in the field.  Several years ago, we published a contribution in the form of a review \cite{doi:10.1142/9789814632683_0001} and the present paper has a different purpose.
It aims at presenting advances in our understanding of scaling above the upper critical dimension 
{over the past hundred years} with the progress and setbacks inherent to the evolution of science.

So, the context is that of second-order phase transitions. Let us {recall} some of the essential results of the theory of phase transitions and critical phenomena. 
According to the standard picture, six critical exponents, $\alpha$, $\beta$, $\gamma$, $\delta$, $\nu$ and $\eta$, to follow nomenclature coined by Fisher\footnote{Fisher  standardised  definitions and formulas for critical point singularities in 1966 \cite{FisherHistorical}.}, describe the singularities of the ``major'' physical quantities when approaching the critical point
{{$h=0,\   t\to 0^\pm, \, t=0,\ h\to 0^\pm $}}, namely:
\begin{eqnarray}
  c_\infty(  t)\simeq\frac{A^\pm}{\alpha}|  t|^{-\alpha}, & &
 \label{eq-1}\
 \\
 m_\infty(  t)\simeq B^- |t|^{\beta},\   t<0,
 && m_\infty(h)\simeq D_{c} |h|^{1/\delta}, \label{eq-2}
 \\
 \chi_\infty(  t)\simeq\Gamma^\pm|  t|^{-\gamma}, & & \label{eq-3}
 \\
 \xi_\infty(  t)\simeq\xi^\pm|  t|^{-\nu}, &&  \label{eq-4}
\end{eqnarray}
and that of the correlation function right  at the critical point,
\be g_\infty(  t=0,h=0,\vac x)\sim\frac{1}{|\vac x|^{d-2+\eta}}. \label{eq-5}
\ee
In these expressions, $t$ and $h$ measure the distance to the critical point, $t=T-T_c$ and $h=H-H_c$ (with usually $H_c=0$).
As stated, the subscript indicates the thermodynamic limit, where the typical size $L$ (hence number of interacting degrees of freedom) tends to infinity. 
The symbols ``$\simeq$" and ``$\sim$" in equations (\ref{eq-1})--(\ref{eq-5}), although widely used in the context of critical phenomena, may require some clarification. 
They are in no way to be understood as the same  as the {equality} sign ``=". 
The first symbol is usually used with the meaning of an equivalence or first approximation between functions, while $\sim$ refers to a similarity  between functions.
We could say that $c_\infty(  t)\simeq\frac{A^\pm}{\alpha} |t|^{-\alpha}$, and $c_\infty(  t)\sim |t|^{-\alpha}$, but each of these are in no way complete --- tradition holds that they imply there is something more.
It is to be understood that upon this {\em singular} behaviour described by the critical exponent $\alpha$, there could be a bunch of regular terms  (e.g. $D_0+D_1 |t|+\dots$). 
It should also be understood that the leading singular behaviour itself may
need to be completed by correction terms called corrections to scaling which may play a role  further away from the transition temperature. 
The expression for $c_\infty(  t)$
then may take a form like\cite{BERCHE20107}

\vspace{2mm}
\hspace{20mm}
\begin{tabular}{llll}
$c_\infty(  t)$ & $=$ & $D^\pm_0+D^\pm_1 |t|+\dots$ & regular background \\
\noalign{\vskip2pt}
&$+$  & $\frac{A^\pm}{\alpha} |t|^{-\alpha}[1 +
    $ & leading singularity \\
\noalign{\vskip2pt}
&& $\qquad\qquad
+\ a_1^\pm |t|^{\omega}
            +a_2^\pm |t|^{2\omega}+\dots$
            & leading corrections  \\
\noalign{\vskip2pt}
&           & $\qquad\qquad+\ {a'}_1 |t|^{\omega'}
        +{a'}_2^\pm |t|^{2\omega'}+\dots$
        & next corrections \\
\noalign{\vskip2pt}
&           & $\qquad\qquad+\ b_1^\pm |t|
        +b_2^\pm |t|^{2}+\dots]$
        & analytic corrections \\
\noalign{\vskip2pt}
&               & $\times\ (-\ln |t|)^{\hat\alpha}\times
    \left(1+
	\dots )\, . \right.$
        & logarithmic corrections.
\end{tabular}

\vspace{2mm} \noindent 
The amplitudes $A^\pm$  and exponent
 $\alpha$ are associated with the leading singularity. The corrections to scaling due to irrelevant 
fields are denoted as  $a_n^\pm$ and $\omega$, ${a'}_n^\pm$ and $\omega'$, \dots.
There might be also analytic corrections associated to non-linearities of the relevant scaling
fields (here $b_n^\pm$). Eventually,  multiplicative  logarithmic corrections
(here $\hat\alpha$) may be observed. These corrections may have different origins and are discussed in
Refs.~\cite{PrivmanEtAl,Kennalog,KennaPRL1,KennaPRL2}. They
can occur in particular right at the upper critical dimension. 

This being said, four scaling relations exist  among the six standard critical exponents,
\bea
&\nu d=2-\alpha, 
\label{oldhyper} \\
&\alpha+2\beta+\gamma=2,  \\
&\beta(\delta -1)=\gamma, \\
&\nu(2-\eta)=\gamma,
\label{oldeta} 
\eea
and, as a consequence, the knowledge of two of them is enough to {derive} all the others.

The first of these  was developed in 1965 by Widom\cite{Wi65a,Wi65b,Gr67} and later   by Kadanoff\cite{Ka66}.
The second was originally proposed in 1963 by Essam and Fisher\cite{EsFi63}, the third a year later by Widom\cite{Wi64} and the fourth by Fisher\cite{Fi64sc}.
The first equality in this list is often referred to as Josephson's scaling relation but Josephson's contribution was actually an associated inequality ($\nu d \ge 2 - \alpha$) derived a couple of years later~\cite{Jo67}. 
Likewise there is an inequality related to the second formula ($2\beta + \gamma  \ge  2 - \alpha$) which was rigorously proven by Rushbrooke in 1963~\cite{Ru63}.
There is also an inequality ($\beta (\delta - 1)  \le \gamma$) associated with the third relation -- proved in 1967 by Griffiths ~\cite{Gr65}.
An inequality associated with Fisher's scaling relation was proved a few years later~\cite{BuGu69,Fi69inequalities}.
The main focus of this paper is the first scaling relation.
Because it involves dimensionality, it is called the {\emph{hyperscaling}} relation.
It is frequently said to {\emph{fail}} in high dimensions where $\alpha$ and $\nu$ adhere to their mean-field values. In Section~\ref{sec7} we will present a new form for it, valid in all dimensions. This new format  necessitates a new critical exponent which will emerge from the scaling exponents as the other do
but in a less obvious manner.

 A fundamental hypothesis in the theory of critical phenomena is 
 called the {\em homogeneity assumption} for the singular part of
the free energy density~\cite{PatashinskiPokrovski,PhysicsPhysiqueFizika.2.263}. It relies entirely on the RG and
 states that $f_\infty^{\rm sing}$ is a generalized homogeneous function of the arguments $t$ and $h$, the relevant thermal and magnetic fields already introduced,  with corresponding RG eigenvalues $   {y_t} $ and $   {y_h} $:
\be
f_\infty^{\rm sing}(  t,h)=b^{-d}f_\infty^{\rm sing}( b^   {{y_t}}  t, b^   {y_h} h).\label{eq-10}
\ee 
Here, $d$ is the space dimension and $b$ is an arbitrary rescaling factor,
 This equation is often rewritten as
\be
f_\infty^{\rm sing}(  t,h)=|t|^{d/y_t}{\mathscr F}^\pm(h|t|^{-y_h/y_t})
\ee
where ${\mathscr F}^\pm(y)=f_\infty^{\rm sing}(\pm 1,y)$ are two
universal functions (above and below the critical temperature), called  scaling functions,  of the unique argument  $y=h|t|^{-y_h/y_t}$ and defined by (\ref{eq-10}) with the choice of rescaling factor $b=|t|^{-1/y_t}$.  
We will use calligraphic upper case letters to denote such scaling functions when the dependence with some variables is omitted if these variables keep zero or constant values.

The homogeneity assumption is complemented by a similar hypothesis concerning the correlation function and the correlation length, 
\bea
g_\infty(\vac x,  t,h)&=&b^{-2x_\phi}g_\infty(b^{-1}\vac x, b^   {y_t}  t, b^   {y_h} h)\nnb\\
&=&|\vac x|^{-2x_\phi}{\mathscr G}_{\vac u}(|\vac x|/|t|^{-1/y_t},|\vac x|/|h|^{-1/y_h}),
\ \hbox{with}\ b=|\vac x|\, .\label{eq-11}\\
\xi_\infty(  t,h)&=&b\xi_\infty( b^   {y_t}  t, b^   {y_h} h)\nnb\\
&=&|t|^{-1/y_t}\Xi^\pm(h|t|^{-y_h/y_t}),\ \hbox{with}\ b=|t|^{-1/y_t}\, ,\label{eq-12}\\
& \textstyle{\rm or} &
|h|^{-1/y_h}\tilde\Xi^\pm(t|h|^{-y_t/y_h}),\ \hbox{with}\ b=|h|^{-1/y_h}\, .\label{eq-12b}
\eea
Here, $\vac u=\vac x/|\vac x|$, but {because} we will only focus on isotropic critical phenomena, this vector subscript in the scaling function of the correlation function  is unnecessary and will be omitted in the rest of the paper.
The {explicit definitions here are for}  the order parameter correlation function and $x_\phi$ is  the corresponding order parameter scaling dimension which we will discuss later. 
A thermal correlation function could be similarly defined.

The critical behaviours of the
magnetization,  susceptibility,  internal energy and   specific heat then follow by
taking the appropriate derivatives of (\ref{eq-10}) wrt $  t$ or $h$,
\begin{eqnarray}
&&m_\infty(t,h)=b^{-d+   {y_h}}m_\infty( b^   {y_t}  t,b^   {y_h} h),\label{eq-1a}\\
&&\chi_\infty(  t,h)= b^{-d+2   {y_h}}\chi_\infty(b^   {y_t}  t,b^   {y_h} h),\label{eq-1c}\\
&&e_\infty(  t,h)=b^{-d+   {y_t}}e_\infty( b^   {y_t}  t,b^   {y_h} h),\label{eq-1d}\\
&&c_\infty(  t,{h})=b^{-d+2   {y_t}}c_\infty(b^   {y_t}  t,b^   {y_h} h)\label{eq-1e}.
\end{eqnarray}

The definition of the critical exponents according to the standard terminology~\cite{FisherHistorical} then
 follows from the elimination of the
$x=b^   {y_t}  t$  or $y=b^   {y_h} h$  dependencies. {\it At the critical temperature}, the choice $b=|h|^{-1/   {y_h}},$ gives 
\begin{eqnarray}
&m_\infty(0,h)\simeq D_{c}^{-1/\delta}|h|^{1/\delta}, 
    &\delta=\frac{   {y_h}}{d-   {y_h}},
    \quad D_{c}^{-1/\delta}=m_\infty(0,1).\label{defBc}
\end{eqnarray}
{ In zero magnetic field,} $b=|  t|^{-1/   {y_t}}$ delivers
\begin{eqnarray}
&m_\infty(  t,0)\simeq B^- |t|^\beta, 
    &\beta=\frac{d-   {y_h}}{   {y_t}}, 
    \quad B^-= m_\infty(-1,0), 
    \    t < 0\, , \label{defB}\\
&\chi_\infty(  t,0)\simeq \Gamma^\pm|  t|^{-\gamma}, 
    &\gamma=\frac{2   {y_h}-d}{   {y_t}},
    \quad\Gamma^\pm = \chi_\infty(\pm 1,0),\label{defGamma}\\
&c_\infty(  t,0)\simeq\frac{A^\pm}\alpha|  t|^{-\alpha}, 
    &\alpha=\frac{2   {y_t}-d}{   {y_t}}, 
    \quad\frac{A^\pm}{\alpha}=c_\infty(\pm 1,0).\label{defA}
    \end{eqnarray}
The same strategy operates in (\ref{eq-12}) to give
\bea
&\xi_\infty(  t,0)\simeq \Xi^\pm|  t|^{-\nu}, &\nu=\frac{1}{   {y_t}},\quad\Xi^\pm=\xi_\infty(\pm 1,0),\label{defNu}
\\
&\xi_\infty(  0,h)\simeq \tilde\Xi^\pm|  h|^{-\nu_c}, &\nu_c=\frac{1}{   {y_h}},\quad\tilde\Xi^\pm=\xi_\infty(0,\pm 1),\label{defNub}
\eea
and in equation (\ref{eq-11}), the choice $b=|\vac x|$, together with the isotropy requirement, leads at criticality $  t=h=0$ to
\be
d-2+\eta=2x_\phi.
\label{defEta}
\ee

It is now clear that the picture in  Eqs.(\ref{eq-1})--(\ref{eq-4}) is not complete. Using  the notations of
Refs.~\cite{Berche_2013}
for the critical amplitudes one expands it to
\begin{eqnarray}
 h=0,\   t\to 0^\pm, &&   t=0,\ h\to 0^\pm, \nnb\\
 f_{\infty}^{\rm sing}(  t,0)\simeq F^\pm|  t|^{2-\alpha}, && f_{\infty}^{\rm sing}(0,h)\simeq F_c|h|^{1+1/\delta}, \\
 m_\infty(  t,0)\simeq B^-|  t|^{\beta},
 && m_\infty(0,h)\simeq D_{c}^{-1/\delta}|h|^{1/\delta}, \label{eq-49}\\
 e_\infty(  t,0)\simeq \frac{A^\pm}{\alpha(1-\alpha)}|  t|^{1-\alpha}, && e_\infty(0,h)\simeq E_c|h|^{\epsilon_c}, \label{eq-50}\\
 \chi_\infty(  t,0)\simeq \Gamma^\pm|  t|^{-\gamma}, && \chi_\infty(0,h)\simeq \Gamma_c|h|^{1/\delta-1}, \label{eq-51}\\
 c_\infty(  t,0)\simeq \frac{A^\pm}{\alpha}|  t|^{-\alpha}, && c_\infty(0,h)\simeq \frac{A_c}{\alpha_c}|h|^{-\alpha_c}, \label{eq-52a}\\
 m_{T\infty}(  t,0)\simeq m^\pm|  t|^{\beta-1},  && m_{T\infty}(0,h)\simeq m_c|h|^{\epsilon_c-1},\\ 
 \xi_\infty(  t,0)\simeq \xi^\pm|  t|^{-\nu}, && \xi_\infty(0,h)\simeq \xi_c|h|^{-\nu_c}. 
\end{eqnarray}
The cross derivative of the free energy density wrt $t$ and $h$, i.e., the magnetocaloric coefficient $m_T$, is often measured in experiments --- since it is more singular that the magnetization itself --- but 
{ its critical exponent} is not an independent quantity.
The standard definitions of some exponent
combinations which  occur above are: $\alpha_c=\alpha/{ (}\beta\delta{ )}, 
\beta_c=\beta/{ (}\beta\delta{ )}$, $\gamma_c=\gamma/{ (}\beta\delta{ )}$,
$\nu_c=(\delta-1)/(2\delta)=$, $\epsilon_c=1-\alpha_c$. 
This makes all together a {collection} of critical exponents, with many redundancies, the most commonly used being the six that we mentioned initially.

Second-order phase transitions may then be classified according to universality classes in which physical systems share the same set of critical exponents. 
These universality classes essentially depend on the space dimensionality and the symmetries of the order parameter. They also depend on the range of interactions, a situation discussed in Section~\ref{secLRI}. A few universality classes  are given in Table~\ref{tab1}.
\begin{table}[t]
 {\small\footnotesize
\begin{tabular}{lllllllll}
\hline
Symmetry &  models & $2d$  & $3d$ & $4d$ & $5d$ & $6d$ & $7d$ &  $\dots$ \\  
\hline
$Z_q$\\
$Z_1$ & Percolation & $2d$ Perco  & $3d$ Perco  & $4d$ Perco & $5d$ Perco  &  $\phi^3+\hbox{logs}$ & $\phi^3$  \\
$q=2$ & Ising,  & $2d$ IM  & $3d$ IM & $\phi^4+\hbox{logs}$ & $\phi^4$ & $\phi^4$ & $\phi^4$ \\
& binary alloys, \\ & \dots\\
$q=3$ & Potts $q=3$,  & $2d\ Z_3$ & 1st & 1st & 1st & 1st & 1st  \\
& surface adsorption,\\ & \dots\\
$q=4$ & Potts $q=4$, & $2d\ Z_4$ & 1st & 1st & 1st & 1st & 1st   \\
&  Ashkin-Teller, \\ 
& surface adsorption,\\ &\dots\\
\hline
$O(N)$\\
$N\to 0$ & SAW, & $2d$ SAW  & $3d$ SAW  &  $\phi^4+\hbox{logs}$ & $\phi^4$ & $\phi^4$& $\phi^4$ \\
$N=2$ & XY,  & KT & $3d\ O(2)$ & $\phi^4+\hbox{logs}$ & $\phi^4$ & $\phi^4$ & $\phi^4$  \\
& superconductivity,\\
& BEC, \dots\\
$N=3$ & Heisenberg, & No order & $3d\ O(3)$ & $\phi^4+\hbox{logs}$ & $\phi^4$ & $\phi^4$ & $\phi^4$  \\
\hline
\end{tabular}
\caption{Examples of universality classes. 1st denotes first-order phase transitions. 
{Perco, IM, SAW and KT stand for percolation, the Ising model, self-avoiding walk and the Kosterlitz-Thouless transition.} Above a certain space dimension, mean-field theory (here denoted as $\phi^n$) describes properly the critical properties. } \label{tab1}
}\end{table}
What we discuss in these lectures describes the situation which holds everywhere in the table where a $\phi^n$ appears.

At this point we have sketched the general scaling picture based on a homogeneity assumption for the free energy, correlation function and correlation length.
The six main critical exponents $\alpha$, $\beta$, $\gamma$, $\delta$, $\nu$ and $\eta$, together with cross-derivative exponents, are each derivable from the scaling dimensions $y_t$, $y_h$ and $x_\phi$ together with the dimensionality $d$. 
There are  four scaling relations, one of which involves $d$.
We have not yet given values of the critical exponents or values of the scaling dimensions.
In the next section we do the former and in the following section we derive the latter in circumstances where order-parameter fluctuations might be expected to be negligible.
The incompatibility of the two will expose the failure identified by Fisher at the start of this section.

\section{Ginzburg-Landau {mean-field} theory}
\label{LMFT}

{Ginzburg-}Landau theory is a mean-field theory (MFT), i.e. one  which does not take into account the critical fluctuations of the order parameter.
It is based on simple assumptions  which essentially boil down to writing the free energy density as a power expansion of the order parameter and its derivatives. 
This can be applied to many different systems.
{A simplified version in which the order parameter is spatially uniform is called the Landau theory.}

Like any theory with a universal ambition, it cannot be accurate in its details for every individual circumstance. 
One can hardly expect a theory that is able to describe extremely different systems (e.g. liquid-gas transition, superconductivity or ferromagnetism) to at the same time provide exact results for each of these systems. 
Nevertheless, being based on minimal hypotheses,  Landau theory  is an instructive approach and is generally qualitatively {correct}.
Furthermore, Landau theory 
{is}  accurate in most of its predictions in high dimensional systems, for which 
fluctuations cease to play the major role. 
{However, the minimal, instructive and phenomenological features of the model are in no way shortcomings or failings. 
In trimming back to bare essentials, the theory explores the roles  of fundamental concepts such as space dimensionality and order-parameter symmetries for universal critical phenomena, without worrying about the quirks of individual physical set-ups.
}

This theory is covered in virtually any course on phase transitions and critical phenomena, so that the following cursory exposition is probably superfluous for most readers. 
But in order to be complete and to further fix the notations, we will sketch the main lines of Ginzburg-Landau theory.
Let us assume that a physical system is described at thermal equilibrium by the set of equations
\bea
&&Z=\int {D}\phi \ \!e^{-F[\phi]},\label{eq-Z31}\\
&&F[\phi]=\int d^dx\ \! f(\phi,\bnabla\phi),\label{Eq-32}\\
&&f(\phi,\bnabla\phi)={\textstyle \frac 12} r\phi^2(\vac x)+
{\textstyle \frac 13}u_3\phi^3(\vac x)
+{\textstyle \frac 14}u_4\phi^4(\vac x)
+{\textstyle \frac 16}u_6\phi^6(\vac x)
-h\phi(\vac x)+{\textstyle \frac 12}|\bnabla\phi|^2.\label{Eq-33}
\eea
Here, the partition function $Z$ is a functional integral over the values of $\phi(\vac x)$, a real scalar field that we  call the matter field to distinguish it from external fields (temperature, magnetic field, \dots).  $F[\phi]$ is the free energy, a functional of the same field,
and
$f$ is a free energy density.
The limitation that our  presentation here is in terms of a scalar matter field (i.e., with $O(1)$ symmetry) is for the simplicity of notation only; it does not have severe consequences for the overall concepts and vector fields would only describe $O(N)$ theories with higher values of $N$. 
We obviously borrow the denomination of matter field from classical field theories for which one would call $F[\phi]$ an action and $f$ a Lagrangian density.

The coefficient $r\sim  t$ and, for our purposes, we take  $u_n\ge 0$ for all $n$, although the study of first-order phase transitions would partially relax  this condition. 
The coefficient of the highest power must anyway be positive to guarantee stability for a finite equilibrium value of the order parameter. 
More generically, $r$ controls the phase transition and must be positive in the phase for which the equilibrium order parameter vanishes and negative in the ordered phase. 
Usually, we think about this parameter in terms of the distance of the temperature to its critical value $T_c$, but $t$ may also be another type of parameter measuring the distance to some critical value (in percolation theory for example,  there is no temperature).
{Again, we frame our discussion it terms of temperature for expediency and without losing this generality.}
An external magnetic field is essentially represented by $h$ and couples linearly to the matter field.

At the mean-field level, the partition function is dominated by the field configurations $\phi_0(\vac x)$ which have the highest weight,
\be
\left.\frac{\delta F}{\delta\phi}\right|_{\phi_0(\vac x)}=0.\label{eq-33}
\ee
This  is the Euler-Lagrange equation and reads as
\be
\frac{\partial f}{\partial\phi}-\bnabla\cdot\frac{\partial f}{\partial (\bnabla\phi)}=0.
\ee

The Ising model belongs to the $\phi^4$ universality class, for which $u_3=0$ and $u_4>0$, hence there is no need for $u_6$. 
This is the standard Ginzburg-Landau-Wilson model. 
Percolation corresponds to $u_3>0$ and the expansion is stopped there. The tricritical point which marks the singular behaviour at the end of a line of first-order phase transitions, as e.g. in the Blume-Capel model for a specific choice of parameters,  is described by $u_3=u_4=0$, $u_6>0$. Note that the gradient term does not have its own coefficient. This is because it is usually absorbed by a rescaling of all other coefficients. This makes this term dimensionless once integrated over space and thus, calling $x_\phi$ the matter-field dimension, one has
$2(x_\phi+1)=d$ which fixes 
\be x_\phi =\frac d2-1.\label{eq-36a}\ee
{Compare this to Eq.(\ref{defEta}), which is nothing more than power counting, and we see that there is no $\eta$ exponent in mean-field theory.
This is why it is termed the \emph{anomalous dimension} --- a non-zero value is a measure of deviation from the simplest theory.
}

Let us consider then a generic model: 
\be
f(\phi,\bnabla\phi)={\textstyle \frac 12} r\phi^2(\vac x)+
{\textstyle \frac 1n}u_n\phi^n(\vac x)
-h\phi(\vac x)+{\textstyle \frac 12}|\bnabla\phi|^2\label{eq-36}
\ee
with $n=3$, 4, 6 respectively corresponding to percolation, $O(N)$ models (including SAW or polymers) and tricriticality.
In an infinite homogeneous system, the gradient term cancels and  equation (\ref{eq-33}) delivers the equation of state
\be
\phi_0(r+u_n\phi_0^{n-2})=h.\label{eq-37}
\ee
In the absence of a magnetic field, $h=0$, the order parameter $m_\infty=\phi_0$ discriminates between the two phases,
\bea
T<T_c, &\quad& m_\infty(  t)=({-r}/{u_n})^\frac{1}{n-2},\label{eq-38}\\
T>T_c, &\quad& m_\infty(  t)=0.\label{eq-39}
\eea
The high-temperature region where the order parameter vanishes is referred to   the ``symmetric phase'', while the low-temperature phase is  called the ``symmetry broken phase''. 
This terminology refers to the study of broken symmetries which occur when cooling down a system from its disordered phase at high temperature.
In equation (\ref{eq-38}), one can read the critical exponent of the order parameter.  Retaining the terminology of magnetic systems we use the symbol $\beta$, 
\be
\beta_{\rm\scriptscriptstyle MFT}=\frac 1{n-2}.\label{eq-41}
\ee
The corresponding amplitude is  $B^-=(u_n)^{-1/(n-2)}$.
At the critical temperature, $r=0$, the same order parameter  has a magnetic field dependence which we extract from  (\ref{eq-37})
\bea
T=T_c,&\quad&m_\infty(h)=({|h|}/{u_n})^{\frac{1}{n-1}}.\label{eq-42}
\eea
The exponent $\delta$ follows from equation (\ref{eq-2}),
\be
\delta_{\rm\scriptscriptstyle MFT}=n-1\label{eq-43}
\ee
together with the amplitude $D_c=(u_n)^{-1/(n-1)}$.
The  susceptibility requires the calculation of the  second derivative of (\ref{eq-37}) wrt $h $ \Rvsd{(or $\chi=\partial\phi_0/\partial h$)}, leading to the equation
\be\chi_\infty=[r+(n-1)u_n\phi_0^{n-2}]^{-1}\ee
from which one gets two expressions, depending on the values of the order parameter (\ref{eq-38}) and (\ref{eq-39}) in each of the two phases:
\bea
T<T_c, &\quad& \chi_\infty(  t)=[{(n-2)(-r)}]^{-1},\\
T>T_c, &\quad& \chi_\infty(  t)= r^{-1}.
\eea
Both expressions lead to the same exponent $\gamma$, as anticipated in the definition in equation (\ref{eq-3}), 
\be
\gamma_{\rm\scriptscriptstyle MFT}=1.
\label{eq-47}
\ee
The associated amplitudes are $\Gamma^-=
(n-2)^{-1}$ and $\Gamma^+=1$.

The specific heat exponent from equation (\ref{eq-1}) requires the second derivative of the free energy wrt $t$. 
The homogeneous system has a free energy given by inserting the equilibrium order parameter  (\ref{eq-38}) and (\ref{eq-39}) in the expansion (\ref{eq-36}), leading to
\bea
T<T_c, &\quad& f_\infty=\left(\frac 1n-\frac 12\right)u_n^{\frac{2}{2-n}}(-r)^{\frac n{n-2}},\label{eq-47bis}\\
T>T_c, &\quad& f_\infty=0.
\eea
The specific heat follows, vanishing above $T_c$,
\bea
T<T_c, &\quad& c_\infty=\frac 1{2-n}u_n^{\frac{2}{2-n}}(-r)^{\frac {4-n}{n-2}},\label{eq-47ter}\\
T>T_c, &\quad& c_\infty=0,
\eea
and a jump appears at the transition, with an exponent  associated with the  low temperature regime (this is specific to the mean-field solution), 
\be
\alpha_{\rm\scriptscriptstyle MFT}=\frac{n-4}{n-2}.\label{eq-52}
\ee
The amplitude in this regime is $(A^-/\alpha_{\rm\scriptscriptstyle MFT})=(u_n)^{-2/(n-2)}/(2-n)$.

For the correlations, one has to go back to equation (\ref{eq-36}) and keep the gradient term. The Euler-Lagrange equation now
leads to the differential equation
\be
r\phi(\vac x)-u_n\phi^{n-1}(\vac x)-\bnabla^2\phi(\vac x)=h.\label{eq-52bis}
\ee
The space dependence of the correlation function corresponds to the order parameter profile when a localized magnetic field
$h_0\delta(\vac x)$
 is applied at the origin.
At criticality, $r=h=0$ and $\phi(\vac x)$ can be considered small enough to neglect the non-linear term. Outside the origin this leads to a Laplace equation, independently of the value of $n$ at this approximation
\be
\bnabla^2\phi(\vac x)= \frac{1}{|\vac x|^{d-1}} \frac d{d|\vac x|} \left(|\vac x|^{d-1}\frac{d\phi(\vac x)}{d|\vac x|} \right)= 0,\ee
where isotropy is assumed. The solution is of the form
\be
g(\vac x)\sim\frac{1}{|\vac x|^{d-2}}
\ee
and is consistent with the value of the correlation function critical exponent
\be
\eta_{\rm\scriptscriptstyle MFT}=0.\label{eq-56}
\ee
This is the standard result of Ornstein-Zernicke theory.
This exponent does not depend on $n$ as we anticipated.
Outside the critical temperature, (say above $T_c$, since we are still neglecting the $\phi^{n-1}$ term in (\ref{eq-52bis}))
 the equation to solve is
\be
r\phi(\vac x)-\bnabla^2\phi(\vac x)=h.\label{eq-56bis}
\ee
The approximation for the term in $\phi^{n-1}(\vac x)$ can be questioned, but it is not an essential one in the present context, since we are now reasoning on dimensional arguments. 
Clearly, one has to introduce length scales,  and the only length scale in the thermodynamic limit is the correlation length.
We therefore identity the correlation lengths $\xi(t,h=0)$ and $\xi(t=0,h)$ as
\be
\xi(t,h=0) \sim (1/r)^{1/2},\qquad 
\xi(t=0,h)\sim (\phi/h)^{1/2}.
\label{eq-xiapprox}
\ee
 With $\phi \sim h^{1/\delta}$, both  exponents of the correlation length follow:
\be
\nu_{\rm\scriptscriptstyle MFT}=1/2,\qquad
\nu_{\rm\scriptscriptstyle c\ \!MFT}=
\frac{\delta - 1}{2\delta} = \frac{n-2}{2(n-1)}.\label{eq-59}
\ee
We assume the same temperature dependence for the correlation length below $T_c$ (this can be made more rigorous, see e.g. Ref.~\cite{cha95}).
We emphasize that $\nu$ does not depend on $n$.

As special cases,  the set of mean-field exponents for the Ising model, percolation and tricriticality universality classes are collected in Table~\ref{tab2}.

Lattice animals correspond to another universality class and deserve a few words here, since they fit in the general picture that we are presenting.
{These are} the connected clusters that we can form on a lattice.
Like polymers, their sizes and shapes obey specific scaling forms.
It happens that they are described  by a Landau expansion with a $\phi^3$ theory and only one external field, say a temperature-like field, coupled to the linear power of the matter field $\phi$.
We therefore  consider
\be
f(\phi,\bnabla\phi)= r\phi(\vac x)+
{\textstyle \frac 1n}w_n\phi^n(\vac x)
+{\textstyle \frac 12}|\bnabla\phi|^2\label{eq-63bis}
\ee
with $n=3$ for lattice animals. 
If we proceed along the same lines as in the previous calculations, we get the order parameter $\phi_0=(-r/w_n)^{1/(n-1)}$ below $T_c$, hence $\beta_{\rm\scriptscriptstyle MFT}=\frac 1{n-1}$, or $1/2$ for lattice animals. 
The susceptibility varies like $\chi\sim1/\phi^{n-2}$ and delivers $\gamma_{\rm\scriptscriptstyle MFT}=\frac {n-2}{n-1}$ and the specific heat has the same exponent, $\alpha_{\rm\scriptscriptstyle MFT}=\frac {n-2}{n-1}$. 
The exponent $\eta$ is not modified wrt the theory of Eq.~(\ref{eq-36}), $\eta_{\rm\scriptscriptstyle MFT}=0$, and dimensional arguments for the correlation length lead to $\xi\sim(\phi/r)^{1/2}$, hence $\nu_{\rm\scriptscriptstyle MFT}=\frac{n-2}{2(n-1)}$. 
{In the absence of another external field in the model, the exponent $\delta$ can be obtained by scaling relations}
and one gets 
$\delta_{\rm\scriptscriptstyle MFT}=(\beta+\gamma)/\beta=n-1$.
 Collecting all exponents (for $n=3$), one has the last row   in Table~\ref{tab2}.

\begin{table}[ht]
\begin{center}
\begin{tabular}{llccccccc}
\hline
Model & $\phi^n$ & $\alpha_{\rm\scriptscriptstyle MFT}$ & $\beta_{\rm\scriptscriptstyle MFT}$ & $\gamma_{\rm\scriptscriptstyle MFT}$ & $\delta_{\rm\scriptscriptstyle MFT}$ &  $\nu_{\rm\scriptscriptstyle MFT}$ & $\eta_{\rm\scriptscriptstyle MFT}$ & $d_{\rm uc}$ \\
\hline
Magnets, SAW & $\phi^4$ & $0$ & ${\textstyle \frac 12}$ & $1$ & $3$ & ${\textstyle \frac 12}\vphantom{{\displaystyle \frac 12}}$ & $0$ & $4$ \\
Percolation & $\phi^3$ & $-1$ & $1$ & $1$ & $2$ & ${\textstyle \frac 12}\vphantom{{\displaystyle \frac 12}}$ & $0$ & $6$ \\
Tricriticality & $\phi^6$ & ${\textstyle \frac 12}$ & ${\textstyle \frac 14}$ & $1$ & $5$ & ${\textstyle \frac 12}\vphantom{{\displaystyle \frac 12}}$ & $0$ & $3$ \\
Lattice animals & $\phi+\phi^3$ & ${\textstyle \frac 12}$ & ${\textstyle \frac 12}$ & ${\textstyle \frac 12}$ & $2$ & ${\textstyle \frac 14}\vphantom{{\displaystyle \frac 14}}$ & $0$ & $8$ \\
\hline
\end{tabular}
\caption{Mean-field exponents for the Ising model, percolation, tricriticality and lattice animals universality classes.}\label{tab2}
\end{center}
\end{table}

We said that Landau theory provides a quantitatively  {valid} description of phase transitions when  order parameter fluctuations can be neglected. 
There is a self-consistent criterion which  shows that  mean-field exponents indeed lead to neglect of fluctuations above a certain space dimension. 
The fluctuations are essentially measured by the susceptibility, which is the space integral of the order parameter correlation function,
\be
\chi\simeq\int d^dx\ \! g(\vac x)\sim |  t|^{-\gamma_{\rm\scriptscriptstyle MFT}}.
\ee
This has to be compared to the square of the magnetization  inside the correlation volume at the same temperature,
\be
\xi^d m_\infty^2\sim|  t|^{-d\nu_{\rm\scriptscriptstyle MFT}+2\beta_{\rm\scriptscriptstyle MFT}}.
\ee
If the first expression is dominated by the second, then order parameter fluctuations are weakened. This happens with mean-field values of the exponents when  
\be
d\ge\frac{2\beta_{\rm\scriptscriptstyle MFT}+\gamma_{\rm\scriptscriptstyle MFT}}{\nu_{\rm\scriptscriptstyle MFT}}.
\ee
Collecting the values of the mean-field exponents in equations (\ref{eq-41}), (\ref{eq-47}) and (\ref{eq-59}), we get
\be 
d\ge d_{\rm uc}= \frac{2n}{n-2}.
\label{ginzy}
\ee
\Rvsd{When this bound is satisfied, and fluctuations can be neglected, $\<\phi^2\>\sim\<\phi\>^2$ and the mean-field theory is valid. }
This is known as the Ginzburg criterion
and we refer to $d_{\rm uc}$ as the upper critical dimension.
The values of $d_{\rm uc}$ are given in Table~\ref{tab2}.

\section{The Gaussian fixed point and its apparent failure to describe critical phenomena above $d_{\rm uc}$}
\label{Gscaling}

Let us come back to the free energy density 
(\ref{Eq-33}) or (\ref{eq-36}) from which we have deduced the matter field scaling dimension \Rvsd{$x_\phi=d/2-1$ given in Eq.}(\ref{eq-36a}). 
Each term has the dimension of a density (per unit volume), 
and from these the scaling dimensions of the external fields  $  t,h,u$  follow (from now on, we {shorten} the notation {to} $u$ for $u_n$ and $y_u$ for $y_{u_n}$),
\bea
&y_  t+2x_\phi = d, &  {y_t}=2, \label{eq-62}\\
&{y_h}+x_\phi = d, &  {y_h}=\frac d2+1,\label{eq-63}\\
&y_{u}+nx_\phi = d, &  y_{u}=\frac d2(2-n)+n 
.\label{eq-64}
\eea
These scaling dimensions, or RG  eigenvalues, 
control the renormalization flow of the three parameters:
\bea
&&  t'=b^{{y_t}}  t,
\label{MPt}\\
&&h'=b^{{y_h}}h,
\label{MPh}\\
&&u'=b^{y_{u}}u.
\label{MPu}
\eea
There is a trivial fixed point at $  t=h=u=0$.
{It is} called the Gaussian Fixed Point (GFP) 
because there the partition function (\ref{eq-Z31}) becomes a Gaussian integral,
\be
Z=\int D\phi\ \!e^{-\int d^dx\ \!\frac 12|\bnabla\phi|^2}.
\ee

With the additional scaling field $u$, the homogeneous form (\ref{eq-10}) becomes 
\be
f_\infty^{\rm sing}(  t,h)=b^{-d}f_\infty^{\rm sing}( b^   {{y_t}}  t, b^   {y_h} h, b^   {y_u} u)\, .\label{eq-10new}
\ee 
The simplicity of this  scaling form is deceptive as it hides subtle phenomena near the critical point as we shall see. 
The counterpart forms for the correlation function and correlation length are
\be
g_\infty^{\rm sing}(\vac x,  t,h)=b^{-2x_\phi}g_\infty^{\rm sing}(b^{-1}\vac x , b^   {{y_t}}  t, b^   {y_h} h, b^   {y_u} u)\label{eq-10new2}
\ee 
and
\be
\xi_\infty^{\rm sing}(  t,h)=b \xi_\infty^{\rm sing}( b^   {{y_t}}  t, b^   {y_h} h, b^   {y_u} u),\label{eq-10new3}
\ee 
respectively.
The temperature and the magnetic field have positive RG eigenvalues
in Eqs.(\ref{eq-62}) and (\ref{eq-62}). We say that these are relevant fields. 
{They tell} us that under a rescaling by a factor $b>0$, the relevant fields grow as 
\be
  t '=b^{2}  t,\quad\hbox{and}\quad h'=b^{\frac d2+1}h.
\ee
Starting from outside the critical point, with either $  t\not=0$ or $h\not=0$ (or both), renormalization brings the system further and further away from criticality
($t=h=u=0$). 
The scaling field $u$ in Eq.(\ref{eq-64}) {also} has a positive RG eigenvalue below a certain value 
\be d_{\rm uc}=\frac {2n}{n-2}\label{eq-70}\ee 
of the space dimension, meaning that $u$ is relevant there. 
{This is, of course,} exactly the value of $d_{\rm uc}$ in
Eq.(\ref{ginzy}).
{However}, $y_{u}<0$ above $d_{\rm uc}$ {and} here $u$  is said to be irrelevant. 
In this case, even if one starts from a non zero initial value of $u$, successive rescalings drive it to zero and leave the system at criticality.
{The negativity of  $y_u$ or irrelevance of $u$ there is another reason why} Landau theory and mean-field exponents provide a {correct} description of  critical properties above $d_{\rm uc}$.
The field $u$ in this situation does not determine the universal quantities which maintain the values of the GFP.
The border line between the two regimes, precisely at $d_{\rm uc}$, {is} the marginal situation where critical singularities are usually accompanied by multiplicative logarithmic divergences.

The question now {arises how the} {above} {\em correct} mean-field exponents 
{(Table~\ref{tab2}) compare} with the predictions drawn from RG at the Gaussian fixed point.
Using the RG eigenvalues (\ref{eq-62}) and (\ref{eq-63}), and the scaling dimension (\ref{eq-36a}), inserted in equations (\ref{defBc})--(\ref{defEta}), one gets the exponents listed in Table \ref{tabGaussian} {below}. 
The scaling dimensions (\ref{eq-62}) and  (\ref{eq-63}) take the same values,  irrespective of the value of $n$ 
in the free energy density expansion, i.e. {they are} the same for all three universality classes --- the Ising model, percolation and tricriticality --- above their respective upper critical dimensions.
{The $n$ dependency is carried only by Eq.}(\ref{eq-64}) {which,  although the field $u$ does not determine Gaussian critical exponents, we also insert in the table}.

\begin{table}[ht]
\begin{center}
\begin{tabular}{lll}
\hline
$y_  t =2\vphantom{{\displaystyle \frac 12}} $& $y_h =\frac d2+1$ & $y_u =d(1-\frac n2)+n$\\
\hline
$\alpha_{\rm\scriptscriptstyle G}=2-\frac d2$ & $\beta_{\rm\scriptscriptstyle G}=\frac{d-2}4\vphantom{{\displaystyle \frac 12}}$ & $\delta_{\rm\scriptscriptstyle G}=\frac{d+2}{d-2}$\\
$\gamma_{\rm\scriptscriptstyle G}=1$ & $\nu_{\rm\scriptscriptstyle G}=\frac 12\vphantom{{\displaystyle \frac 12}}$ & $\eta_{\rm\scriptscriptstyle G}=0$\\
\hline
\end{tabular}
\caption{{RG eigenvalues and} critical exponents at the Gaussian fixed point. 
{From Eq.(\ref{eq-70}), $y_u = n(d_{\rm uc}-d)/d_{\rm uc} < 0$ if $d > d_{\rm uc}$ so, in contrast to $y_t$ and $y_h$, is irrelevant there. }}\label{tabGaussian}
\end{center}
\end{table}

{ The mismatch between Table~\ref{tab2} and 
Table~\ref{tabGaussian} for some of the exponents is obvious!}
While the third row of Gaussian exponents in Table~\ref{tabGaussian} do indeed coincide with mean-field exponents, the agreement is broken by the second row. 
{This is clearly a major 
``flaw in the original [RG] argument'' for the RG}, since, as we have argued earlier, mean-field exponents are correct above $d_{\rm uc}$, where $u$ is irrelevant. 
{This is the ``failure'' that Fisher was referring to in the second quote of Section~\ref{sec1} above.}

In trying to understand the origin of the failure of the GFP above $d_{\rm uc}$, we notice that the {(wrong)} exponents $\alpha$, $\beta$ and $\delta$ all come from derivatives of the free energy, while {the (right)} $\gamma$, $\nu$ and $\eta$ are associated with the correlations. 
We will refer to the first set of quantities as belonging to the ``free energy sector'' and the second group as belonging to the ``correlation sector''. 
The susceptibility is special in the sense that it is at the same time associated with a free energy derivative and  the correlation function integral. Still, an explanation of the discrepancy {is needed for} the free energy sector.
{(We will see that it is also needed for the apparently unproblematic correlation sector too where extremely subtle incompatibilities hide.)}
Incidentally, we can also observe that {all exponents, including the obviously} ``deviant'' {ones} coincide with  the mean-field counterparts precisely at the respective values of $d_{\rm uc}$ for the three universality classes discussed here. In addition, the GFP value for 
$\nu_{\rm\scriptscriptstyle c}$
is $\nu_{\rm\scriptscriptstyle c\ \! G} = 2/(d+2)$ from inserting Eq.(\ref{eq-63}) for $y_h$ into 
Eq.(\ref{defNub}), and this does not agree with the mean-field value in Eq.(\ref{eq-59}) except, again, when $d=d_{\rm uc}$. This is the sign that something is wrong also in the correlation sector!

\section{The Dangerous Irrelevant Variable scenario}
\label{DIVscene}
\subsection{{Fisher's breakthrough}}
\label{breakthrough}

In 1983,\footnote{Although the proceedings were published in 1983, the lecture itself was given in 1982. The idea probably germinated  in Fisher's mind a lot earlier as another lecture dating from 1973 is often cited in this context~\cite{Gunton1973RenormalizationGI}.  } revisiting the question of discordance between the Gaussian fixed point and mean-field theory, Michael Fisher\cite{FisherStellenbosch} made a very smart observation. 
Although one would expect the GFP to deliver the correct predictions above $d_{\rm uc}$ ({where} the exponents should not depend on $u$), Fisher noticed that for the three quantities in the ``free energy sector'' that lead to the exponents $\alpha$, $\beta$ and $\delta$, the limit $u\to 0$ in the expressions obtained in Landau theory is problematic. 
This comes from the amplitudes in equations (\ref{eq-38}), (\ref{eq-42}) and (\ref{eq-47ter}) which are singular when $u\to 0$, 
\bea
&&B^-=u^{-\frac 1{n-2}},\label{eq_76}\\
&&D_c=u^{-\frac 1{n-1}},\label{eq_77}\\
&&\frac{A^-}{\alpha_{\rm\scriptscriptstyle MFT}}=\frac1{2-n}u^{-\frac 2{n-2}}.\label{eq_78}
\eea
The irrelevant  field $u$ is therefore  {\em dangerous} and we speak about the role of the {\em dangerous irrelevant variable} (DIV).
{Not addressing these is the ``failure to recognize and allow for possible singular behaviour of the scaling function'' that Fisher was referring to in the quote of Sec.\ref{sec1}.}
The amplitudes of the quantities in the ``correlation sector'', on the other hand, do not depend on $u$ and do not face the same difficulty (with the notable exception of $\xi_c$, but the $h-$dependence of the correlation length at $T_c$ is rarely discussed, so we also leave this quantity aside). 
Another argument given by Fisher and Privman\cite{Privman1983} is that a strictly positive value of $u$ is required to ensure the stability of the free energy. 

Here we {introduce} the following notation {to compactify} the {exponents} appearing in (\ref{eq_76})--(\ref{eq_78}): 
\bea
 &&m_\infty(  t<0,h=0,u)\sim  |t|^\beta u^{-\kappa}, \\
 &&m_\infty(  t=0,h,u)\sim |h|^{1/\delta} u^{-\lambda},\\
 && c_\infty(  t,h=0,u)\sim |  t|^{-\alpha}
u^{-\mu},
\eea
{where in} $\phi^n$  Landau theory, these parameters take the  values
\be
\kappa=\frac 1{n-2},\quad\lambda=\frac 1{n-1},\quad\mu=\frac 2{n-2}.\label{E81}
\ee
{These} are collected
 in Table~\ref{tab4} for the universality classes under consideration.

\begin{table}[ht]
\begin{center}
\begin{tabular}{lllll}
\hline
Model & $\phi^n$ & $\kappa$ & $\lambda$ & $\mu$  \\
\hline
Magnets, SAW & $\phi^4$ & $\frac 12\vphantom{{\displaystyle \frac 12}}$ & $\frac 13$ & $1$  \\
Percolation & $\phi^3$ & $1$ & $\frac 12\vphantom{{\displaystyle \frac 12}}$ & $2$ \\
Tricriticality & $\phi^6$ & $\frac 14\vphantom{{\displaystyle \frac 12}}$ & $\frac 15$ & $\frac 12$  \\
\hline
\end{tabular}
\caption{Exponents of the dangerous irrelevant variable in Landau theory.}\label{tab4}
\end{center}
\end{table}

To accommodate these observations, equations  (\ref{eq-1a}) and (\ref{eq-1e}) have to be modified, taking into account the dependence on the DIV $u$, and its dangerous limit of $u\to 0$.
E.g., to draw the singularity in the scaling function {for the magnetisation} to the fore we express it as 
\be 
m_\infty(x,0,z)\stackrel{u\to 0}{=}z^{-\kappa} {\mathscr M}^-(x,0) .
\ee 
The scaling hypothesis for the magnetization and specific heat are now modified and must obey 
\bea
&&m_\infty(  t,0,u)
\stackrel{u\to 0}{=}b^{-d+{y_h}-\kappa y_{u}}u^{-\kappa}{\mathscr M}^-(b^{y_t}t,0),\label{eq-81}\\
&&m_\infty(0,h,u)\stackrel{u\to 0}{=}b^{-d+{y_h}-\lambda y_{u}}u^{-\lambda}{\mathscr M}_c(0,b^{y_h}h),\\
&&c_\infty(  t,0,u)\stackrel{u\to 0}{=}b^{-d+2{y_t}-\mu y_{u}}u^{-\mu}{\mathscr C}^\pm(b^{y_t}t,0) \label{eq-85}.
\eea
On the other hand, nothing has to be modified for the ``correlation sector'' (susceptibility, correlation function, correlation length).
Fixing the scaling factor $b$ to the appropriate value, $|  t|^{-1/   {y_t}}$ or 
$|h|^{-1/   {y_h}}$ in (\ref{eq-81})--(\ref{eq-85}), then leads,
{to}
\bea
\beta	_{\rm\scriptscriptstyle MFT}=\frac{d-   {y_h}}{   {y_t}}+\frac{\kappa y_u}{   {y_t}}
={\frac{1}{n-2}},\label{E85-2}
\\
\frac{1}{\delta_{\rm\scriptscriptstyle MFT}}
=\frac{d-   {y_h}}{   {y_h}}+\frac{\lambda y_u}{   {y_h}}
={\frac{1}{n-1}},\label{E85-1}
\\
\alpha_{\rm\scriptscriptstyle MFT}
=\frac{2   {y_t}-d}{   {y_t}}-\frac{\mu y_u}{   {y_t}}
={\frac{n-4}{n-2}}
\label{E85}
\eea
which completely repairs the free energy sector above $d_{\rm uc}$. 
{The values of $\kappa$, $\lambda$ and $\mu$ in Eq.(\ref{E81}) are precisely those that match Eqs.(\ref{eq-41}), (\ref{eq-43}) and (\ref{eq-52}), which are gathered here for convenience.}
The amplitudes also should be consistent and we find for example for the magnetization approaching the critical temperature from below 
\be
B^-={\mathscr M}^-(0^-) u^{-\kappa}.
\ee

We have now reached a point where RG appears {to be} fully successful in its treatment of critical properties  above the upper critical dimension in the thermodynamic limit {at least}. 
{To refer again to the quote in Sec.~\ref{sec1}, ``the renormalization group framework has been preserved intact'' --- or so it seems.} 
{However, besides for notational purposes in the prelude (Sec.~\ref{SecNotations}), we have so far not touched on finite-size systems.
We address finite-size scaling in the next section and encounter another conflict that again raises question about  RG.
The dichotomy this time is between predictions coming from MFT and  analytical results from Br\'ezin et al.~\cite{Brezin82} as well as numerical results from Binder et al. \cite{Binder85}, all dating from the 1980's.
We present these in the next section along with partial solutions up to the 1990's.
}

\subsection{The problem with finite-size scaling}
\label{probwFSS}

In  the standard theory of finite-size scaling (FSS), 
if,  in the thermodynamic limit, a physical quantity $Q(t,h)$ 
is described by a critical exponent $\rho$ wrt temperature, say,
i.e., \be Q_\infty(  t,0)\sim |  t|^\rho,
\label{noplace}
\ee
one usually takes that its finite-size counterpart $Q_L(  t,0)$ at a given temperature $t$ is controlled by the ratio of the finite size $L$ to the typical length scale which governs criticality, the correlation length $\xi_\infty(  t)$ at the same temperature,
\be
Q_L(  t,0)=Q_\infty(  t,0)f_Q(L/\xi_\infty(  t)) \label{eq86}.
\ee
The finite system cannot deliver any singularity because the partition function is a finite sum of exponentials and can only display regular behaviour.
Therefore  we demand \Rvsd{for $x$ large} that the function $f_Q(x)\sim x^\omega$  corrects the singularity in $Q_\infty$.
This in turn implies that $\omega=-\rho/\nu$, and we obtain the FSS behaviour 
$ Q_L(  t,0)=A_Q(  t)L^{-\rho/\nu}$. 
To get rid of the temperature dependent prefactor, one usually fixes $  t=0$ to
{sit at the critical point and} obtain
\be Q_L(  t=0,0)\sim L^{-\rho/\nu}.\label{eq87}\ee
The argument is also encoded in the scaling hypothesis and is probably more convincing there. Using  the case of the susceptibility  for example, 
 \Rvsd{extending} equation (\ref{eq-1c}) \Rvsd{to systems of finite sizes requires the introduction of a new scaling field $L^{-1}$ which resacles as $bL^{-1}$, then  the choice $b=L$ and $h=0$} leads to
\be
\chi_L(  t,0)
=L^{\frac{\gamma}{\nu}}{\mathscr X}(L^2  t).\label{eq-88}
\ee
At the critical point $  t=0$ this {gives} 
$\chi_L(t=0)
\sim L^{\frac{\gamma}{\nu}}$, but usually FSS is performed at the pseudo-critical point $  t_L$ instead. {This} is easier to determine {in finite-size} numerical {simulations} than the true critical temperature {which requires extrapolation to} the thermodynamic limit. The pseudo-critical temperature $T_L$ can be defined by, e.g., the value of the temperature for which the finite-size susceptibility (or any other diverging quantity) reaches its maximum, $\chi_L(T_L)=\hbox{Max}_T\ \!\chi_L(T)$, a quantity often denoted in the literature as $\chi_{\rm max}(L)$, but that  we will call here $\chi_L(  t_L,0)$  with $  t_L=T_L-T_c$, the second argument being the magnetic field, as usual.
There, an expansion of (\ref{eq-88}) leads to
$\chi_L(  t_L,0)\simeq L^{\frac{\gamma}{\nu}}{\mathscr X}(t=0)+\dots$ and thus, up to corrections to scaling, to the same leading FSS behaviour as at $T_c$.

Inserting MFT exponents \Rvsd{(\ref{E85-2})-(\ref{E85}), and $\nu=1/2$} in equation~(\ref{eq87}), one thus expects 
the FSS behaviour
\bea
&&{c_L(t=0,0)\sim L^{2\frac{n-4}{n-2}}},\\
&&\chi_L(t=0,0)\sim L^2,\\
&&m_L(t=0,0)\sim L^{-{\frac{2}{n-2}}},\\
&&\xi_L(t=0,0)\sim L. \label{Eq96}
\eea
We call this {\emph{Landau FSS}} because the exponents which appear in powers of $L$ are all ratios of exponents from Landau theory.

These predictions, however, fail.
The first theoretical analysis of  finite-size correlation length for the $\phi^4$ model above $d_{\rm uc}=4$ dimensions was reported in the early 1980's  by Br\'ezin~\cite{Brezin82}, who obtained
\be \xi_L(t=0,0)\sim L^{d/4}\label{EQ_95}\ee
which contradicts Landau scaling (\ref{Eq96}). 
Br\'ezin had considered  hypercubic systems with periodic boundary conditions. 
The same author, with Zinn-Justin, then {(1985)} produced a {more} complete study of FSS in phase transitions in the $\phi^4$ model~\cite{BREZIN1985867} and reported for example the susceptibility behaviour, above $d_{\rm uc}$,
\be \chi_L(t=0,0)\sim L^{d/2},\label{EQ_97}\ee
with the comment that ``usual FSS does not hold''. These authors  conclude their paper with 
\begin{quotation}
{\em It seems clear that in spite of  the extensive literature on the subject, there is still a lot to say about finite-size effects.}
\end{quotation}

In early numerical studies {(1985)}, Binder reported  results for the finite-size susceptibility of the  Ising model in 5 dimensions with PBC's~\cite{Binder85} (see also \cite{Rickwardt}). 
In particular, at the pseudo-critical point he obtained
\be \chi_L(  t_L,0)\sim L^{5/2},\quad\hbox{and}\quad    t_L\sim L^{-5/2}.
\label{E95}
\ee
Binder  also led a discussion of the specific heat maximum there, but we believe that due  the available sizes being too small  ($L\le 7$), the conclusions were probably not sound
(in such early studies the thermal sector was not fully understood).
The relation (\ref{EQ_95}) of Br\'ezin was later checked numerically   by Jones and Young~\cite{PhysRevB.71.174438}.

Later, other quantities have been calculated via {MC} simulations by many people (see the review~\cite{BERCHE2012115} and references therein), and, if we quote only the results for the $5d$ Ising model with PBC's, we collect 
\bea
&\hbox{pseudo-critical point}\quad &\hbox{critical point}\nnb\\
\hbox{correlation length}&\xi_L(  t_L,0)\sim L^{5/4},\quad &\xi_L(t=0,0)\sim L^{5/4},\\
\hbox{susceptibility}&\chi_L(  t_L,0)\sim L^{5/2},\quad &\chi_L(t=0,0)\sim L^{5/2},\\
\hbox{magnetization}&m_L(  t_L,0)\sim L^{-5/4},\quad &m_L(t=0,0)\sim L^{-5/4},\\
\hbox{pseudo-critical temperature}&  t_L\sim L^{-5/2},\\
\hbox{rounding of the critical point}&\Delta\beta_{\chi_{\rm max}/2}\sim L^{-5/2},\\
\hbox{pseudo-critical magnetic field}&|h_L|\sim L^{-15/4},\\
\hbox{Lee-Yang zero}&h^{\rm LY}_L(  t_L)\sim L^{-15/4},&h_L^{\rm LY}(t=0)\sim L^{-15/4},\\
\hbox{Fisher zero}&&t^{\rm F}_L(  h=0)\sim L^{-5/2}.
\eea
We have discussed the first four rows and the other quantities require some explanation.

The rounding $\Delta\beta_{\chi_{\rm max}/2}$ is the width of the temperature window as measured at half the susceptibility height. The quantity denoted as $|h_L|$ 
is the shift of magnetic field at the critical temperature, i.e. the finite value of the magnetic field 
($\pm |h_L|$) at which the susceptibility peaks at $t=0$. 
{(This does not occur in zero magnetic field).} 
Finally, $h^{\rm LY}$ is the first Lee-Yang zero and $t^{\rm F}$ the first Fisher zero, to be discussed in Section~\ref{seczeros}.

There is clearly a disagreement \Rvsd{between the measured Finite-Size Scaling} above $d_{\rm uc}$ with Landau FSS, and it does not seem to be solved by Fisher's DIV mechanism \Rvsd{which only repairs the values of the exponents in the thermodynamic limit}. 
{Even a desperate attempt to invoke the (wrong) GFP values for the critical exponents of \Rvsd{Table~\ref{tabGaussian}} fails to rescue the situation; while they deliver different FSS in the free-energy sector (namely, $c_\infty \sim L^{\alpha_G/\nu_G}=L^{4-d}$ and $m_\infty \sim L^{-\beta_G/\nu_G}= L^{-(d-2)/2}$), they deliver the same as Landau FSS for $\chi$ and $\xi$. And, as we have seen, these are $n$-independent and not in agreement with exact or numerical results for the Ising case of $n=4$.}
This calls for further developments and this is the motivation of Binder, Nauenberg, Privman and Young's (BNPY) approach \cite{PhysRevB.31.1498}.

\subsection{Dangerous irrelevancy for the free energy and the thermodynamic length }
\label{DIVFSS}

In 1985, Binder, Nauenberg, Privman and Young  suggested an extension of the DIV mechanism\cite{PhysRevB.31.1498}, but  we believe that it {was}  not  fully developed.
It results in a theory compatible with  Fisher's {and, indeed, Landau MFT,} and involves an additional hypothesis {that at least partially} solves the problem of FSS. 
We first present this approach and the way in which finite-size effects are understood in BNPY's theory and then we expose what we believe are still weak points calling for {further} extensions {of Fisher's DIV concept}.

We have seen with Fisher's analysis that the homogeneity assumption has to be modified to take into account the existence of the DIV $u$. 
BNPY suggested to build on Fisher's suggestion to reconsider, e.g., the magnetization and the specific heat,  
assuming that not only the prefactors, 
but also the rescaled  arguments of the free energy might be affected.
The modification propagates to the dimension of  prefactors in the free energy derivatives. 
 They made the following hypothesis in the dangerous limit:
\be 
f_\infty^{\rm sing}(x,y,z)\stackrel{{z}\to 0}{=}z^{p_1}f_\infty^{\rm sing}(xz^{p_2},yz^{p_3},0)
\label{BNPYL1}
\ee
which
{ can be}
rewritten in a more convenient form {[cf Eq.(\ref{eq-10})]:}
\be
f_\infty^{\rm sing}(  t,h,u)\stackrel{u\to 0}{=}b^{-d+p_1y_u}u^{p_1}{\mathscr F}^\pm( b^{   {y_t}+p_2y_u}  tu^{p_2}, b^{   {y_h}+p_3y_u} hu^{p_3}).\label{eq-102}
\ee 
{Here} $p_1$, $p_2$ and $p_3$ are constants which have to be determined by further considerations.
BNPY also introduced the notations
\be d^*=d-p_1y_u,\quad
y_  t^*=   {y_t}+p_2y_u,\quad y_h^*=   {y_h}+p_3y_u,\label{E104}
\ee
and also assumed that the correlation length could obey a similar homogeneity law,
\be
\xi_\infty(  t,h,u)\stackrel{{u\to 0}}{=}b^{1+q_1y_u}u^{q_1}{\Xi}( b^{   {y_t}+q_2y_u}  tu^{q_2}, b^{   {y_h}+q_3y_u} h u^{q_3}),
\label{eq-104}
\ee 
with {the} three other parameters,  $q_1$, $q_2$ and $q_3$, at that point unknown.

Binder and his coauthors  then developed an argumentation to support the values (they presented their results for $n=4$ but we generalise them here
\bea
&&p_1=0,\label{eq-105}\\
&&q_1=0.
\eea
The first result follows from the assumption $d^*=d$,  and is underpinned by a discussion on the zero-field susceptibility for a finite system. The second result is supported as follows: 
\begin{quotation}{\em Since the finite-size correlation length $\xi_L$ is bounded by $L$, we require $q_1 y_u<0$. {\rm (}\dots{\rm )} if one adopts the plausible assumption that for $  t=h=0$, the correlation length increases up to the linear dimensions of the lattice, which implies that $q_1=0$.}
\end{quotation}
We will see later that the value $q_1=0$ is not correct.

The values of $p_2$ and $p_3$ are not explicitly written in BNPY's paper, but an immediate consequence of the work reported for the $\phi^4$ case {considered there} is  
\be 
 p_2=-{\textstyle\frac 12},
 \quad p_3=-{\textstyle\frac 14}.
  \quad {\text{(These formulae are for $n=4$ only.)}}
\ee
In particular, BNPY discussed the scaling 
with the values $y_  t^*=d/2$ and $y_h^*=3d/4$ {(for $n=4$, again)}  which fix $p_2$ and $p_3$. 
On the other hand, they did not pursue a discussion of $q_2$ and $q_3$, leaving the option that these parameters might differ from $p_2$ and $p_3$, an opinion that we do not share, as 
{we discuss} later.

Although this is not done in the original BNPY paper, we believe it is instructive to explicitly state the complete agreement with Fisher's scenario. 
Here we do this in the context of the $\phi^n$ model of equation (\ref{eq-36}).
Using the appropriate derivatives of the free energy density wrt $t$, and to $h$, we get the first four ``classical'' exponents in the BNPY approach
\bea
&\displaystyle \alpha_{\rm\scriptscriptstyle MFT}=\frac{2   {{y_t^*}}-d}{   {{y_t^*}}}, 
&\beta_{\rm\scriptscriptstyle MFT}=\frac{d-   {{y_h^*}}}{   {{y_t^*}}},
\label{SH1}\\
&\displaystyle \gamma_{\rm\scriptscriptstyle MFT}=\frac{2  {{y_h^*}} -d}{  {{y_t^*}}},
&\delta_{\rm\scriptscriptstyle MFT}=\frac{   {{y_h^*}}}{d-   {{y_h^*}}}.
\label{SH2}
\eea
The comparison with the results of Fisher, (\ref{E85}), leads to the following results for the $p_i$'s parameters (Table~\ref{tab5}),
\be
{p_1=0}, \quad  p_2=-\frac{2\kappa   {y_t}}{d+2\kappa y_u} {=-\frac{2}{n}}, \quad p_3=-\frac{\kappa(2   {y_h}-d)}{d+2\kappa y_u} {=-\frac{1}{n}},\label{E114}
\ee 
having used Eq.(\ref{E81}) for the general $n$ case.
Inserting these values in Eq.(\ref{E104}) gives
\be 
d^*=d,\quad
y_ t^*=   {y_t}-\frac{2y_u}{n},\quad y_h^*=   {y_h}-\frac{y_u}{n},
\label{E1041}
\ee
In terms of $d$ and $n$ these scaling dimensions are
\be 
y_t^* = \frac{d(n-2)}{n}, \quad
y_h^* = \frac{d(n-1)}{n}. \label{E1041b}
\ee
Inserting $y_t^*$ for $y_t$ and $y_h^*$ for $y_h$ in (\ref{defBc}) (\ref{defB}) (\ref{defGamma}) 
in (\ref{defA})
delivers the correct Landau MFT critical exponents for $\alpha$, $\beta$, $\delta$ and $\gamma$ in (\ref{eq-41}), (\ref{eq-43}), and (\ref{eq-47}), (\ref{eq-52}).
The remaining main critical exponents $\nu$ and $\eta$ are unaffected because the correlation sector is not touched upon in Eq.(\ref{eq-102}).

\begin{table}[ht]
\begin{center}
\begin{tabular}{llcccc}
\hline
Model & $\phi^n$ & $p_2$ & $p_3$ & $y_  t^*$ & $y_h^*$  \\
\hline
Magnets, SAW & $\phi^4$ & $-\frac 12\vphantom{{\displaystyle \frac 12}}$ & $-\frac 14$ & $\frac d2$ & $\frac{3d}4$  \\
Percolation & $\phi^3$ & $-\frac 23$ & $-\frac 13\vphantom{{\displaystyle \frac 12}}$ & $\frac d3$ & $\frac{2d}3$ \\
Tricriticality & $\phi^6$ & $-\frac 13\vphantom{{\displaystyle \frac 12}}$ & $-\frac 16$ & $\frac{2d}3$ & $\frac{5d}6$ \\
{General} & {$\phi^n$ }& {$-\frac{2}{n}\vphantom{{\displaystyle -\frac{2}{n} }}$} & {$-\frac{1}{n}$} & {$(1-\frac{2}{n})d$} & {$(1-\frac{2}{n})d$ }\\
\hline
\end{tabular}
\caption{Renormalization of the RG eigenvalues by the dangerous irrelevant variable in BNPY theory, extended here to percolation theory and to tricriticality.}\label{tab5}
\end{center}
\end{table}

Thus far we have formulated Fisher's DIV concept for the free energy itself (in the thermodynamic limit). We have still not touched the (seemingly unbroken) correlation sector.
What is probably the most important assumption made in BNPY paper, as well as in Ref.\cite{Binder85}, concerns FSS.
The authors were aware of the failure above $d_{\rm uc}$ of the ordinary scenario which works well below $d_{\rm uc}$ and proposed to repair it via the introduction of a new length scale, the  \emph{thermodynamic length} $l(  t,h)$ which, instead of the correlation length, would be the relevant scale which controls finite-size effects there. Various reasons in favor of the thermodynamic length were given and in particular the fact that such a length scale, $l\sim|  t|^{-1/y_  t^*},$ appears in (\ref{eq-102}) when the first argument is written in the dimensionless form $(L/l)^{y_  t^*}$.
With this new hypothesis to hand,  equation (\ref{eq86})  {would} be replaced, above $d_{\rm uc}$ by
\be
Q_L(  t,0)=Q_\infty(  t,0)f_Q(L/l(  t))
\ee
leading to the FSS behaviour
\be
Q_L(  t=0,0)\sim L^{-\rho y_  t^*}
\ee
instead of (\ref{eq87}), and in the  particular case of  the susceptibility for the Ising model, to
\be
\chi_L(  t=0,0)\sim L^{\gamma y_  t^*}= L^{d/2}
\ee
where  $\gamma=\gamma_{\rm\scriptscriptstyle MFT}=1$ and $y_  t^*=d/2$ are used. This result  conforms to the numerical simulations of Binder~\cite{Binder85}. These simulations are also consistent with $\gamma=\gamma_{\rm\scriptscriptstyle MFT}=1$.

{At this point the mismatch between RG and analytical/numerical results for the Ising susceptibility appears fixed.} 
We have to stress that in their paper, BNPY did not discuss the FSS behaviour of the correlation length, but their result, $q_1=0$, is incompatible with the calculation of Br\'ezin~\cite{Brezin82} (see also \cite{Caracciolo01}), as we will discuss later in this paper. 
We believe that the main reason for this wrong result, which was pretty common {even up to recent times}, is probably due to an erroneous belief  concerning the very nature of the correlation length. This length is not a {\em material} length. This is an abstract  length defined as the typical  scale of an exponential decay. From this point of view, there is no need to demand physical limitations like the system size. 
In our opinion, if one considers that a DIV affects the free energy density, possibly modifying the usual $b^{-d}$ prefactor, one can just as well contemplate that it affects the correlation length, possibly changing its standard $b$ prefactor. 
In fact, BNPY showed that the $b^{-d}$ prefactor in the free-energy case is unchanged.
This may  influence one's expectation concerning the correlation length, but we believe that the modification of the correlation length prefactor is still as legitimate as the one of the magnetization, or of the susceptibility for example. 
{We address that in Section~\ref{sec7}.}

\subsection{Dangerous irrelevancy and finite-size corrections}
\label{sec6.4}

Erik Luijten and Henk Bl\"ote~\cite{PhysRevLett.76.1557,PhysRevLett.76.3662.3}, using a RG analysis,  provided a wonderful explanation of Fisher and BNPY's DIV scenarios, which otherwise seem ad-hoc in nature.
{{In his  doctoral thesis, a tour de force developed under the direction of  Bl\"ote, Luijten made a sound analysis of the $4$D and $5$D Ising models, including with Long-range Interactions, above their upper critical dimensions~\cite{PhHLuijten}.}}
The approach of Luijten and Bl\"ote provides a more complete mechanism, and  almost closes the question concerning finite-size-scaling above the upper critical dimension {at least for the free energy sector}. 
In their original paper, they considered the $O(N)$ $\phi^4$ model and here we extend the results  to the $\phi^n$ model of equation (\ref{eq-36}).
The study  is based on RG equations for the scaling fields which, limited to linear order, are of the form\footnote{In Ref.~\cite{PhHLuijten}, the parameter $p$ is denoted $3ac$, and since  Luijten and Bl\"ote consider the long-range interaction model (discussed later), $y_  t=\sigma$ and $y_u=2\sigma-d=\epsilon$.} 
\bea
&&\frac{dr}{d\ln b}=   {y_t} r+
	p u,\label{eq115}\\
&&\frac{du}{d\ln b}=y_u u,\label{eq116}\\
&&\frac{dh}{d\ln b}=   {y_h} h.\label{eq117}
\eea
Integration of these equations leads to  
\bea
&&u'=b^{y_u}u,\\
&&r'=b^{   {y_t}}\left(
r-\frac{p}{(y_u-   {y_t})}u+\frac{p}{(y_u-   {y_t})}b^{y_u-   {y_t}}u
\right),\label{eq-119}\\
&&h'=b^{{y_h}}h.\label{eq-120}
\eea
The details of the integration with the inclusion of non linear terms on the RHS of  (\ref{eq116}) and (\ref{eq115}) can be found in Refs.~\cite{PhHLuijten,flore2016}. 
The notation is simplified via the introduction of a constant
\be
\tilde p=-\frac{p}{(y_u-   {y_t})}
\label{clip}
\ee
and the identification of the temperature scaling field,
\be
  t=r+\tilde p u
\ee
leading to the  expression
\be
r'=b^   {y_t}(  t-b^{y_u-   {y_t}}\tilde p u ).
\ee

Now,  we define the second moment of the order parameter
\be
\langle \phi^2 \rangle =\frac{
\int d\phi \ \! \phi^2 \ \! e^{ -F[\phi] }
}
{
\int d\phi \ \! e^{-F[\phi]}
}
\ee
where, for  homogeneous systems, $F[\phi]=L^d(\frac 12r\phi^2+\frac 1n u\phi^n-h\phi)$. A change of variable
\be
\phi=u^{-\frac 1n}\varphi
\ee
absorbs the DIV and
leads to 
$F[\varphi]=L^d(\frac 12r^*\varphi^2+\frac 1n \varphi^n-h^*\varphi)$
where 
\be r^*=r u^{-\frac{2}{n}}\quad\hbox{and}\quad h^*=h u^{-\frac{1}{n}}\ee and is such that 
\be
\langle \phi^2 \rangle=u^{-\frac 2n}\langle\varphi^2\rangle.
\ee
Equations (\ref{eq-119}) and (\ref{eq-120}) are then transposed to a rescaling of $r^*$ and $h^*$,
\bea
{r^*}'&=&r'u'^{-\frac 2n}\nnb\\
&=&b^{   {y_t}-\frac 2n y_u}(  t-b^{y_u-   {y_t}}\tilde p u )u^{-\frac 2n},\label{eq-127}\\
{h^*}'&=&h'u'^{-\frac 1n}\nnb\\
&=&b^{   {y_h}-\frac 1n y_u}h u^{-\frac{1}{n}}.\label{eq-128}
\eea

This expression shows how the DIV contaminates the temperature and magnetic field, and modifies accordingly the homogeneity assumption for the singular part of  the free energy density, including also the system size as an additional scaling field. It reads now as
\be
f_L(  t,h,u)=L^{-d}{\mathscr F}^\pm[
L^{y_  t^*}(  t u^{-2/n}-\tilde p u^{(n-2)/n}L^{y_u-   {y_t}}),
L^{y_h^*}hu^{-1/n}
]\label{E122}
\ee
{with $y_t^*$ and $y_t^*$ defined in Eq.(\ref{E1041}.)}
In (\ref{E122}), $L$ is the finite linear scale of the sample (say the length of the edge of a hypercube), and  the free energy density is no longer singular.

In our opinion, {the analytic inclusion of corrections to scaling in} equation (\ref{E122}) is really a central result.
{The two-term structure was already proposed in BNPY for FBC's but not for PBC's.
While they proposed that the free energy scales as Eq.(\ref{BNPYL1}) (with $b=L$) for PBC's, ``for other boundary conditions, where the system has a surface, it is probably necessary to use both $tL^{y_y^*} $ and $tL^{1/\nu}$ for a complete asymptotic description''~\cite{PhysRevB.31.1498}. 
The second term in Eq.(\ref{eq-133}) when $n=4$  
corresponds to BNPY's proposal and extends it to other boundary conditions, including PBC's. }

Eq.(\ref{E122}) contains essentially FSS as a natural consequence. 
Luijten and Bl\"ote for example  demonstrate the size dependence of the shift of the pseudo-critical temperature in the following way.
If we set the first argument of  (\ref{E122}) scaled by $L$ as
\be
X=L^{y_  t^*}(  t u^{-2/n}-\tilde p u^{(n-2)/n}L^{y_u-   {y_t}}),
\label{E1611}
\ee
 we can denote by $X_0$ the value taken by this variable when a diverging quantity, say the susceptibility, reaches its maximum value wrt the temperature for a finite system. This temperature is, by definition,  the pseudo-critical temperature $  t_L$, which thus obeys the following scaling
\be
  t_L=X_0u^{2/n}L^{-y_  t^*}+\tilde p uL^{y_u-y_  t}.\label{eq-133}
\ee
If the first term dominates, which will always occur for some large $L$, because $y_  t-y_u\ge y_  t^*$ above $d_{\rm uc}$, we recover, in the case of the Ising model, the FSS of Binder in (\ref{E95}) for $d=5$.

At this point the homogeneity assumption for the free energy sector has been modified and the size-dependency of the pseudocritical temperature has entered the game. 
We will later develop the theory of Luijten and Bl\"ote but we present now three additional elements which were studied more recently.

\section{Zeros of the partition function}\label{seczeros}
\label{zerosec}

We {return to the partition function (\ref{ZZZ}) of Sec.\ref{SecNotations} and, e.g., for the Ising model} define it  as a sum over the configurations of the microscopic degrees of freedom $\{s_i\}$: 
\be
Z=\sum_{\{s_i\}}e^{\beta J\sum_{(i,j)} s_is_j+\beta H\sum_i s_i}.
\ee
The partition function encodes the thermodynamic properties through its relation to the free energy $F=-\beta^{-1}\ln Z$, and when the latter  becomes singular at a phase transition, the former approaches zero. 
The zeros in the complex magnetic-field plane are called Lee-Yang zeros\cite{PhysRev.87.404,PhysRev.87.410} and those in the complex temperature field are called Fisher zeros\cite{FisherZeros}.
Lee-Yang and Fisher zeros are, in a sense, the most 
fundamental quantities in terms of which most 
of the thermodynamics quantities can be defined ---
so much so that they form the basis of what is referred 
to as the ``fundamental theory of phase transitions" \cite{Wu}.
Any exposition of the fundamentals of RG and scaling theory should incorporate them and that is the aim of this section.

We  discuss first the Lee-Yang zeros.
For convenience, we keep the Ising model as an example and we define new degrees of freedom $\sigma_i=\frac 12(1+s_i)$ which take the values $0$ and $1$. The partition function  of a finite system with 
$N=L^d$ sites
is rewritten as a sum over microstates with defined values of the energy $E=-\sum_{(i,j)} s_is_j$ ($E$ can take positive and negative integer values ranging from $-dN$ and $+dN$) and the magnetization $S=\sum_i s_i=2\sum_i\sigma_i-N=2M-N$ (where $M= \sum_i\sigma_i$ takes positive integer values from $0$ to $N$).
Then, the partition function reads as
\bea 
Z_L(\beta_c)&=&\sum_{M=0}^{N}\sum_{E=-dN}^{dN}p(E,M)e^{-\beta_cHN}e^{-\beta_c (E+2HM)}
=e^{-\beta_cHN}\sum_{M=0}^N g_M z^{2M}\label{e152}\eea
with the fugacity $z=e^{-\beta_cH}$, and the $g_M$'s are positive numbers (degeneracies of the microstates) which do not depend on $H$. 
One thus observes that
equation (\ref{e152}) is a $N$-th order polynomial in $z^2$ with positive coefficients. 
As a consequence, $Z_L(\beta_c)=0$ is an equation which has only complex roots $z^{ (k)}$. This is consistent with the fact that, 
for a finite system, the free energy 
is analytic at any real value of the magnetic field $H$, hence the partition function has no zero for real $H$. 
Therefore, if we allow for complex values of the magnetic field, $z=e^{g+{\rm i}h}$ ($g,h\in\mathbb R$),
$Z_L(\beta_c)$ has $N$ non-real zeros $z^{ (k)}$ in the complex plane. This allows us to factorize
\bea
&&Z_L(\beta_c)=A(z)\prod_{k=1}^N(z-z^{ (k)}_L(\beta_c)),\label{e153}\\
&&z^{ (k)}=r^{ (k)}e^{{\rm i}\phi^{ (k)}},\quad\hbox{with}\quad r^{ (k)}=e^{-2\beta_c\Re(H)}=e^{g^{ (k)}},\ \phi^{ (k)}=-2\beta_c\Im(H)=h^{ (k)}
\eea
with $A(z)$ a smooth non-vanishing function and $\Im (H_n)\not=0$.
The Lee-Yang theorem states that the zeros lie on the unit circle of the variable $z$, i.e. $z^{ (k)}=e^{{\rm i}\phi^{ (k)}}$ or, along the imaginary axis of the variable $H$~\cite{PhysRev.87.404}.

When the system size increases,  the order of the  polynomial increases, and  with it the  number of zeros.
In the thermodynamic limit,  $L\to\infty$, the zeros on the imaginary axis of the variable $g+{\rm i} h$ become dense.
There is a gap between the real magnetic-field axis and a certain point ${\rm i}h^{\rm LY}(  t)$ for $ \beta<\beta_c$.
That point is called the Lee-Yang edge.
 A phase transition occurs in zero magnetic field at $\beta_c$, which means that
 the gap vanishes when $\beta$ approaches the critical point $  \beta=\beta_c$.
 The vanishing of the gap is controlled by a power law,  involving the so-called gap exponent $\Delta$, so that $h^{\rm LY}(  t)\sim   |t|^\Delta$. 
 The finite-size scaling of the Lee-Yang edge follows from ordinary scaling: the rescaled arguments of the free energy density are $b^{y_  t}  t$ and $b^{{y_h}}h$, and, with the choice $b=  |t|^{-1/y_  t}$, they become $1$ and $  |t|^{-{y_h}/y_  t}h$, showing that the correct scaling between $h$ and $  t$ is indeed
\be
h^{\rm LY}(  t)\sim   |t|^\Delta,\quad\hbox{where}\quad \Delta=
{y_h}/y_  t=\beta\delta=\beta+\gamma.\label{e155}
\ee
 The expression (\ref{e155}) can be written as $h^{\rm LY}(  t)\sim   \xi^{-\Delta/\nu}$ which, for a finite system below the upper critical dimension translates into
$h^{\rm LY}_L\sim   L^{-\Delta/\nu}$. 
For finite systems, Itzykson et al~\cite{ITZYKSON1983415} showed that zeros approximately scale at the critical temperature, according to their rank $k$, as \be h^{ (k)}\sim(k/L^d)^{\Delta/\nu d}.\label{156}
\ee
From equation (\ref{e153}), we get
\bea
&&f_{L}(  0,h)=-L^{-d}\sum_k\ln(z-z^{ (k)}_L( \beta_c)),\\
&&\chi_L(  0,h)=-L^{-d}\sum_k\frac {1}{(z-z^{ (k)}_L(\beta_c))^2}\to_{H\to 0} N^{-1}\sum_{k=1}^N(1-z^{ (k)}_L(\beta_c))^{-2}
\label{eq-158}
\eea
where the last term on the right of (\ref{eq-158}) is evaluated in zero magnetic field. It follows that
\be
\chi_L(t=0,0)\simeq {\rm const}\times N^{-1}\sum_{k=1}^N(h^{ (k)}_L(\beta_c))^{-2}
\simeq N^{2\Delta/d\nu-1}
\ee
where the sum is dominated by the lowest zeros.
This leads to $\chi_L\sim L^{(2\Delta-d\nu)/\nu}$ which is consistent with the ordinary FSS behaviour of the susceptibility, $\chi_L\sim L^{\gamma/\nu}$, when we make use of the scaling relations $\Delta=\beta+\gamma$, $d\nu=2-\alpha$ and
$\alpha+2\beta+\gamma=2$.

Above the upper critical dimension, 
{the Landau exponents of Section~\ref{LMFT} suggest}
\be\Delta_{\rm\scriptscriptstyle MFT}=\frac {n-1}{n-2}\ee
{for the mean-field gap exponent. This}
is $\Delta_{\rm\scriptscriptstyle MFT}=\frac 32$ for the Ising  universality class.
However, we now know that the scaling 
 $\chi_L\sim L^{\gamma/\nu}$ with MFT exponents is not correct, and above $d_{\rm uc}$, we  have to {make appropriate adjustments.}
 FSS {now} suggests
\be
h^{ (k)}_L\sim
L^{-y_h^*}\label{E157}
\ee
from where the susceptibility  follows as
\be
\chi_L(t=0,0)\sim 
L^{2y_h^*-d}.
\ee
The scaling form (\ref{E157}) for the Lee-Yang edge has been checked for the Ising model in $d=5$ dimensions 
in Ref.~\cite{BERCHE2012115}.

The Fisher zeros at $H=0$ in the complex temperature plane are analyzed in a similar manner. 
They  have been studied in Flores-Sola's doctoral thesis which is openly available at~\cite{flore2016}. 
There, using obvious notations,  
one expects above the upper critical dimension FSS to be
\be
  t^{ (k)}_L\sim L^{-y_  t^*}.\label{E159}
\ee
This scaling form  has been confirmed numerically in the case of the long-range Ising model discussed later. This  is consistent with the FSS of the specific heat 
\be
c_L(t=0,0)\sim L^{2y_t^*-d}
.
\ee

\section{Scaling of the Fourier modes }
\label{modes}

The Ginzburg-Landau-Wilson action (\ref{Eq-32}), with the Lagrangian density of the $\phi^4$ theory, namely
\be
F_{\rm GLW}[\phi(\vac x)]=\int d^dx \Bigl({\textstyle \frac 12} r\phi^2(\vac x)+
{\textstyle \frac 14}u\phi^4(\vac x)
-h\phi(\vac x)+{\textstyle \frac 12}|\bnabla\phi|^2\Bigr),\label{E-159}
\ee
can be expressed in Fourier space using the propagating modes $\psi_{\vac k}$ in hypercubic systems with PBC's,
\be
\phi(\vac x)=\sum_{\vac k}\tilde\phi_{\vac k}\psi_{\vac k}=\frac{1}{\sqrt V}\sum_{\vac k}\tilde\phi_{\vac k} e^{i\vac k\cdot\vac x},\quad \vac k=\frac{2\pi}{L}\vac n,\ \vac n\in\mathbb Z^d\label{E-160}
\ee
as
\be
F_{\rm GLW}[\tilde\phi_{\vac k}]=\half\sum_{\vac k}(r+|\vac k|^2)|\tilde\phi_{\vac k}|^2+
\quarter uL^{-d}\sum_{\vac k_1,\vac k_2,\vac k_3}\tilde\phi_{\vac k_1}\tilde\phi_{\vac k_2}
\tilde\phi_{\vac k_3}
\tilde\phi_{-(\vac k_1+\vac k_2+\vac k_3)}-
hL^{d/2}\tilde\phi_0 \, . \label{E161}
\ee
It is useful to consider the zero mode $\tilde\phi_0$ separately to the non-zero modes $\tilde\phi_{\vac k\not=0}$.
The former contributes to the non-vanishing average order parameter in the ordered phase.
The latter do not develop non-zero average values even below the critical temperature (see Section~\ref{SecNotations}). 
Therefore we write an expansion with essentially two different types of terms: {depending or not on the zero mode,}
\bea
F_{\rm GLW}[\tilde\phi_0,\tilde\phi_{\vac k\not=0}]&\simeq& \half \Bigl(r
+\frac{3u}{2L^d}\sum_{\vac k\not=0}|\tilde\phi_{\vac k}|^2
\Bigr)\tilde\phi_0^2+\quarter\frac{u}{L^d}\tilde\phi_0^4-{h}{L^{d/2}}\tilde\phi_0\nnb\\&&\quad+
\half\sum_{\vac k\not=0}(r+|\vac k|^2)|\tilde\phi_{\vac k}|^2+\dots\label{E162}
\eea
where we have not explicitly written terms higher than Gaussian for the non zero modes. 
One can clearly see how the DIV $u$ contaminates 
the temperature field $r$  to quadratic order in the zero mode expansion while no contamination of the same type  appears for the non-zero modes (second line in equation~(\ref{E162})).
The behaviour of $\tilde\phi_0$ is thus controlled by the GFP modified by the presence of the dangerous irrelevant variable, while the $\tilde\phi_{\vac k\not=0}$ modes are governed solely by the GFP.

 We may therefore infer that the anomalous scaling (which we refer to as Q scaling), like in equations~(\ref{EQ_95}) or (\ref{EQ_97}) (this is presented in more detail in Section~\ref{sec7}),
  does not hold for the non-zero modes.  
 Wittmann and Young~\cite{PhysRevE.90.062137} analyzed several $\vac k\not=0$ modes and concluded that the susceptibilities calculated from these  modes indeed display
 {\em standard} FSS  in 
\be
\chi_{\vac k\not=0}
=L^d\langle|\tilde\phi_{\vac k\not=0}|^2\rangle_L
\sim L^2\label{E163}
\ee
with 
\be
\tilde\phi_{\vac k}
=\frac{1}{\sqrt V} \int d^dx\ \!\phi(\vac x)\ \!e^{-i\vac k\cdot\vac x} .
\ee
We nevertheless believe that this statement may be  misleading, since labeling the power of $2$ in equation~(\ref{E163}) as {\em standard} FSS   can either refer to Landau scaling, or to the Gaussian Fixed point scaling (see Table~\ref{tab6}).
The common belief was that {\em standard} FSS is Landau scaling. 
Indeed, as far as we know, we were the first  to refer explicitly to Gaussian FSS for equation~(\ref{E163}), but like others, our understanding of the problem has evolved with time\footnote{In our first
contribution in this topic \cite{BERCHE2012115}, we referred to the $\chi_L\sim L^2$ behaviour as {\em Gaussian} scaling, having in mind MFT or Landau scaling.} so it is an important matter to be able to distinguish the two situations.
 
An obvious strategy to discriminate between both scenarios\footnote{Note that both scenarios emerge directly from the action (\ref{E-159}) and are therefore implicit in the homogeneous form (\ref{eq-10new2}). This single form, along with corrections to be discussed in Sec.\ref{sec6.4} are sufficient to describe scaling and FSS. } is to analyse the FSS behaviour of the ``non-zero modes magnetization'' for example. Indeed, we predict for this quantity
 a Gaussian FSS 
\be
m_{\vac k\not=0}=
\langle|\tilde\phi_{\vac k\not=0}|\rangle_L
\sim L^{-\frac {d-2}2}\label{E164}
\ee
instead of $L^{-1}$ for Landau scaling, and we have checked in Ref.~\cite{PhysRevLett.116.115701} that (\ref{E164}) is indeed correct.
{We refer to the GFP scaling described in Section~\ref{Gscaling}, which manifests as Eqs~(\ref{E163}) and (\ref{E164}),  as G scaling in order to distinguish it from the Q scaling associated with zero modes and from standard FSS that comes from the Landau picture in Section~\ref{LMFT}.}

\section{The Long-range Ising model}\label{secLRI}
The long-range  Ising model (LRIM) is a variant of the Ising model in which the value of the upper critical dimension can be controlled by the exponent which governs the decay of the spin-spin interactions. This has   been extensively studied in the context of FSS above the upper critical dimension by Luijten and Bl\"ote~\cite{PhysRevLett.76.3662.3,PhysRevB.56.8945,PhHLuijten}.

The LRIM is defined by the Hamiltonian
\be
{\mathscr H}=-\sum_{i,j}J(\vac x_i-\vac x_j) s_{\vac x_i}s_{\vac x_j}-H\sum_i s_{\vac x_i}
\ee
where, as for the ordinary Ising model, the spins, located at positions $\vac x_i$, take the values $s_{\vac x_i}=\pm 1$,  but the exchange interaction is governed by the algebraic decay
\be
J(\vac x_i-\vac x_j)=\frac{J}{|\vac x_i-\vac x_j|^{d+\sigma}}.\label{E173}
\ee 
The parameter $\sigma$ is free and $\sigma\ge 2$ gives the same universality class as the nearest neighbour model for which 
$J(\vac x_i-\vac x_j)=J\delta (\vac x_i-\vac x_j-{\pmb\mu})$ where 
${\pmb\mu}$ are the generating vectors of the unit cell.
In Fourier space, the counterpart of the expansion (\ref{E162}) is
\bea
F_{\rm GLW}[\tilde\phi_0,\tilde\phi_{\vac k\not=0}]&\simeq& \half \Bigl(r
+\frac{3u}{2L^d}\sum_{\vac k\not=0}|\tilde\phi_{\vac k}|^2
\Bigr)\tilde\phi_0^2+\quarter\frac{u}{L^d}\tilde\phi_0^4-{h}{L^{d/2}}\tilde\phi_0\nnb\\&&\quad+
\half\sum_{\vac k\not=0}(r+|\vac k|^2+C_\sigma|\vac k|^\sigma)|\tilde\phi_{\vac k}|^2+\dots\label{E174}
\eea
where the  $C_\sigma$  term arises from the Fourier transform of the interaction term in (\ref{E173}).

The RG analysis of the model leads to the GFP anomalous dimensions reported in Table~\ref{tabGFPLRI}.
\begin{table}[ht]
\begin{center}
\begin{tabular}{llll}
\hline
$x_\phi=\frac{d-\sigma}{2}$ &
$y_  t =\sigma\vphantom{{\displaystyle \frac 12}} $& $y_h =\frac {d+\sigma}{2}$ & $y_u =2\sigma -d$\\
\hline
$\alpha_{\rm\scriptscriptstyle G}=2-\frac d\sigma$ & $\beta_{\rm\scriptscriptstyle G}=\frac{d-\sigma}{2\sigma}\vphantom{{\displaystyle \frac 12}}$ & $\delta_{\rm\scriptscriptstyle G}=\frac{d+\sigma}{d-\sigma}$ & $d_{\rm uc}=2\sigma$ \\
$\gamma_{\rm\scriptscriptstyle G}=1$ & $\nu_{\rm\scriptscriptstyle G}=\frac 1\sigma\vphantom{{\displaystyle \frac 12}}$ & $\eta_{\rm\scriptscriptstyle G}=2+\sigma$\\
\hline
\end{tabular}
\caption{Critical exponents at the Gaussian fixed point of the LRIM. The value $\sigma = 2$ gives  the ordinary nearest neighbour Ising results ($\phi^4$).}\label{tabGFPLRI}
\end{center}
\end{table}
The modified scaling dimensions are obtained following the same lines as in equations (\ref{eq-127}) and (\ref{eq-128}),
\bea
y_  t^*=y_  t(\sigma)-\half y_u(\sigma)=\frac d2,\\
y_h^*={y_h}(\sigma)-\quarter y_u(\sigma)=\frac {3d}4,
\eea
and we find the same values as for the case of  the nearest-neighbour model.
It is clear also that the first line in (\ref{E174}), 
which controls the zero mode, is the same as in the short-range model.

Our predictions for the FSS behaviour of zero and non-zero modes (we limit ourselves to the magnetization and susceptibility)  are thus the following: 
\bea
&\hbox{Q scaling:}&m_{0}=
\langle|\phi_{0}|\rangle_L
\sim L^{-\frac {d}4},\\
&&\chi_{0}
=L^d\langle|\phi_{0}|^{2}\rangle_L
\sim L^{\frac d2},\\
&\hbox{G scaling:}&m_{\vac k\not=0}=
\langle|\tilde\phi_{\vac k\not=0}|\rangle_L
\sim L^{-\frac {d-\sigma}2},\\
&&\chi_{\vac k\not=0}
=L^d\langle|\tilde\phi_{\vac k\not=0}|^{2}\rangle_L
\sim L^{\sigma}.
\eea
These expressions have been carefully checked in Refs.~\cite{PhysRevLett.116.115701} and \cite{flore2016}, where LRIM with values of $\sigma=0.1$ or $0.2$ were studied for $1$D and $2$D  (i.e. in the regime above the upper critical dimension $d_{\rm uc}(\sigma)=2\sigma$). 
Once again the behaviour of the non-zero modes clearly appears to differ from Landau standard scaling, and is governed by the pure GFP exponents instead.

\section{Dangerous irrelevancy in the correlation sector}
\label{sec7}

Since Fisher's breakthrough forty years ago rescued the RG framework for the free-energy sector in the thermodynamic limit (Sec.\ref{breakthrough}) above the upper critical dimension, DIVs were extended to finite-size scaling 
corrections (Sec.\ref{sec6.4}).
However, problems still remain 
and in this section we further build on Fisher's legacy to preserve the RG framework.
In particular, we extend DIV's to the correlation sector.
{In doing so,  closely follow} Luijten and Bl\"ote's approach that we  proposed to generalize  in Ref.~\cite{PhysRevLett.116.115701}

\subsection{The use of \coppa, the rescue of hyperscaling and finite-size scaling}

We return firstly to Section~\ref{DIVFSS} and BNPY's extension of the DIV mechanism to the free energy itself.
As stated, they assumed that the correlation length could obey a similar homogeneity law (\ref{eq-104}) but insisted that $q_1=0$ and left $q_2$ and $q_3$ open. 
For the free energy in Eq.(\ref{eq-102}), BNPY gave three reasons why $p_1=0$ and hence $d^*=d$. 
In a series of papers \cite{CMP2013} we advocated that the correlation length should be allowed exceed the length --- as Br\'{e}zin had shown for $d=4$ and general $n$.
In particular, we showed that 
\be
 \xi_L \sim L^{\qqq} 
\ee
with \Rvsd{the exponent ``koppa''}
\be
\mathcoppa=d/d_{\rm uc}.
\label{coppadoverdc}
\ee
Therefore we write
\be
\xi_L(t,h,u)=b^{\qqq}\Xi^\pm(b^{y_t^{+}}tu^{q_2},b^{y_h^{+}}hu^{q_3}).
\label{Fri1}
\ee
where 
\be 
\mathcoppa= 1+q_1y_y,\quad
y_ t^+ =   {y_t}+ q_2 y_u,\quad y_h^+ =   {y_h}+q_3 y_u\, .
\label{Fri2}
\ee
The value for $\mathcoppa$ in Eq.(\ref{coppadoverdc}) determines $q_1=-1/n$.
In the thermal sector, matching Eq.(\ref{Fri1}) with the value $\nu_{\rm\scriptscriptstyle MFT} = 1/2$ in Eq.(\ref{eq-59}) determines 
$q_2=-2/n$, identical to $p_2$. 
Likewise, matching Eq.(\ref{Fri1}) with the value $\nu_{\rm\scriptscriptstyle c\ \!MFT} = (n-2)/2(n-1)$ in Eq.(\ref{eq-59})  determines 
$q_3=-1/n$, identical to $p_3$.
Note that, while the correct correlation-length critical exponent $\nu_{\rm\scriptscriptstyle MFT}$ is identical to that coming from the GFP 
$\nu_{\rm\scriptscriptstyle GFP} = 1/y_t=1/2$ in Eq.(\ref{defNu}), their counterparts in the magnetic sector do not coincide and
$\nu_{\rm\scriptscriptstyle c\ \!MFT}$ differs from
$\nu_{\rm\scriptscriptstyle c\ \!GFP} = 1/y_h$ from Eq.(\ref{defNub}).
So, technically although $\nu_{\rm\scriptscriptstyle MFT} = 1/y_t$, this is by coincidence only and the correct expression is $\nu_{\rm\scriptscriptstyle MFT} = 1/y_t^+ = 1/y_t^*$. 
Note also that because $q_2=p_2$ and $q_3=p_3$, one has $y_t^+=y_t^*$ and $y_h^+=y_h^*$. We henceforth only use the starred scaling dimensions.
I.e., to extend Eqs.(\ref{SH1}) and (\ref{SH2}) to the correlation sector, we write
\bea
&\displaystyle \nu_{\rm\scriptscriptstyle MFT}=\frac{   {{\qq}}}{   {{y_t^*}}} 
&\nu_{\rm\scriptscriptstyle c\ \!MFT}=\frac{\qq}{   {{y_h^*}}}.
\label{SH3}
\eea
Equations (\ref{E1041}) and (\ref{E1041b}) can now be written above $d_{\rm uc}$ in a very consistent manner as
\be
d^*=\qq d_{\rm uc},\quad y_t^*(d)=\qq y_t(d_{\rm uc}),\quad y_h^*(d)=\qq y_h(d_{\rm uc}).
\ee

Likewise, we believe that there is no reason to invoke a mechanism by which the rescaling of the temperature and the magnetic field in equations (\ref{eq-127}) and (\ref{eq-128}) would differ for different physical quantities. A second extension proposed is to use the correct (starred) scaling dimension ($2x_\phi^*$ in fact) 
of the matter field to describe the dimension of the correlation function. 
From these considerations, we are led to the following scaling hypotheses (equation (\ref{E122}) is rewritten for the sake of clarity),
\bea&&f_L(  t,h,u)=b^{-d}{\mathscr F}^\pm(X,Y,bL^{-1}),\label{eq-134}\\
&&g_L(\vac x,  t,h,u)=b^{-2x_\phi^*}{\mathscr G}^\pm(b^{-1}\vac x,
X,Y,bL^{-1}),\label{eq-135}\\
&&\xi_L(  t,h,u)=b^{\qqq}\Xi^\pm(X,Y,bL^{-1}).\label{eq-136}
\eea
These are expressed in terms of the two rescaled variables
\bea
&&X=b^{y_  t^*}(  t u^{-2/n}-\tilde p u^{(n-2)/n}b^{y_u-   {y_t}}),\label{eq-137}\\
&&Y=b^{y_h^*}hu^{-1/n},\label{eq-138}\eea
and with the scaling dimensions
\be
y_  t^*=d\Bigl(1-\frac 2n\Bigr),\ y_h^*=d\Bigl(1-\frac 1n\Bigr),\ x_\phi^*=\frac dn,\ \mathcoppa=
\frac d2\Bigl(1-\frac 2n\Bigr)\label{eq-139}
\ee
having used 
(\ref{E1041}) with (\ref{eq-62}), (\ref{eq-63}), (\ref{eq-64}) and extended $x_\phi^*$ from Ref.~\cite{PhysRevLett.116.115701} for the $n=4$ case, and where
$\tilde p$ is given in Eq.(\ref{clip}).
{Obviously these scaling forms apply to the  Q-sector only (Fourier $Q$-modes). 
The G sector is unaffected by DIVs and the finite-size counterparts of Eqs.(\ref{eq-10}), (\ref{eq-11}) and (\ref{eq-12}) apply there.  }
Following a suggestion by Michael Fisher, a  new exponent  $\mathcoppa$  was introduced in \cite{CMP2013} {for the  Q sector}.\footnote{We are indebted to Michael Fisher who, having invited us to meet him in the Royal Society just before he retired in 2012, suggested the usage of this nice archaic Greek letter instead of the letter $q$ {which we had used hitherto} \cite{BERCHE2012115}.
The notation $q$ itself came from usage of $\hat{q}$ for the exponent governing the logarithmic correction to scaling of the correlation length in \cite{KennaPRL1,KennaPRL2} and as a nod to BNPY's usage of $q_1$ in Eq.(\ref{eq-104}) \cite{PhysRevB.31.1498}.} We like the use of this exponent, since it is very easy to translate equations from one universality class to another in terms of $d/d_{\rm uc}$, and also to generalize from the GFP values (which we recover reverting $\qq$ to 1), so one can also rewrite the exponents (\ref{eq-139}) in the form
\be
y_t^*=2\mathcoppa,\ y_h^*=\frac d2+\mathcoppa,\ x_\phi^*=\frac d2-\mathcoppa,\ \mathcoppa=\frac d{d_{\rm uc}}.
\ee

An interesting use of $\mathcoppa$ is in a new form of the hyperscaling relation.
Setting $b=|t|^{-1/y_t}$ in (\ref{eq-136}) delivers $\xi_\infty\sim|t|^{-\qqq/y_t^*}$, hence $y_t^*=\mathcoppa/\nu$, then in (\ref{eq-134}) we get $f_\infty\sim|t|^{d/y_t^*}\sim|t|^{2-\alpha}$ which leads to 
\be
\alpha=2-\frac{\nu d}{\qqq}.
\label{hyperhyper}
\ee
This repairs\footnote{Since 
Josephson's inequality $\nu d \ge 2 - \alpha$ was introduced in 1967 \cite{Josephson67}, {literature, including} textbooks, on statistical physics, lattice field theory, etc. refer to  hyperscaling as ``failing" above the upper critical dimension.  
This statement should no longer be used in statistical physics --- hyperscaling does not fail because the RG does not fail above the upper critical dimension. 
{Moreover, the}  hyperscaling relation should rather be rewritten properly {as (\ref{hyperhyper}) and not as (\ref{oldhyper}).} } the hyperscaling relation above $d_{\rm uc}$, and it holds also below $d_{\rm uc}$ where $\mathcoppa=1$.
{(Obviously $\mathcoppa=1$ for the non-zero or orthogonal Fourier modes in high dimensions too.) }

The new exponent already enabled predictions to be more naturally expressed  for models such as the nearest-neighbour Ising model, {percolation} above its critical dimension $d_{\rm uc}=6$  and for LRIM's with various dimensions above $d_{\rm uc}(\sigma)$  (with periodic boundary conditions)
\cite{lotsfq,Deger2020}.
The extension to general values of $n$ is obvious and  we collect the predictions for arbitrary $d>d_{\rm uc}$ for quantities which have been discussed above:
\bea
&&m_L\sim L^{-(\frac d2-\qqq)},\ \chi_L\sim L^{2\qqq},\ e_L\sim L^{-(d-2\qqq)},\ c_L\sim L^{4\qqq-d},\label{eq-140}\\
&&\xi_L\sim L^{\qqq},\ g_L(X_0)\sim L^{-(d-2\qqq)},\label{eq-141}\\
&&  t_L\sim L^{-2\qqq},\ \Delta\beta_L\sim L^{-2\qqq},\ |h^L|\sim L^{-(\frac d2+\qqq)},\\
&&h^{\rm LY}_L\sim L^{-(\frac d2+\qqq)}, \ t^{\rm F}_L\sim L^{-2\qqq}.\label{eq-202}
\eea
Note that all of the above scaling formulas are readily obtained by replacing the standard FSS prescription that converts Eq.(\ref{noplace}) to Eq.(\ref{eq87}) by the prescription
\be Q_\infty(  t,0)\sim |  t|^\rho
\longrightarrow 
 Q_L(  t=0,0)\sim L^{-\qqq \rho/\nu}.
 \label{QFSS}
 \ee
 This is what we call Q-scaling and it holds at the pseudocritical point as we shall see shortly.
 With Q-scaling to hand, FSS holds above the upper critical dimension.

 The different approaches that we have discussed so far are collected and compared in Table~\ref{tab6} for the $\phi^4$ model with periodic boundary conditions. 
 In this table, the first column lists known results in the thermodynamic limit or for finite-size scaling. 
 The remaining columns correspond to the different approaches that we have described so far, with the symbol $\surd$ to denote an agreement and, {in cases of disagreement}, the prediction made by the (incorrect) theory considered is given explicitly. The last column is for Q scaling. 

\begin{table}[t]
\begin{center}
\begin{tabular}{llllll}
\hline
The correct results & Landau scaling$^1$ & GFP$^2$ & Fisher DIV$^3$ & BNPY$^4$ & \coppa$^5$ \\
\hline 
$c_\infty(  t,0)\sim|  t|^0$ & $\surd$ & $\alpha_{\rm\scriptscriptstyle G}=2-\frac d2$ &  $\surd$ & $\surd$ & $\surd$ \\
$m_\infty(  t,0)\sim|  t|^{1/2}$ & $\surd$ & $\beta_{\rm\scriptscriptstyle G}=\frac{d-2}4$ &  $\surd$ & $\surd$ & $\surd$  \\
$\chi_\infty(  t,0)\sim|  t|^{-1}$ & $\surd$ & $\surd$ &  $\surd$ & $\surd$ & $\surd$ \\
$m_\infty(0,h)\sim|h|^{1/3}$ & $\surd$ & $\delta_{\rm\scriptscriptstyle G}= \frac{d+2}{d-2}$ &  $\surd$ & $\surd$ & $\surd$ \\
$\xi_\infty(  t,0)\sim|  t|^{-1/2}$ & $\surd$ & $\surd$ &  $\surd$ & ? & $\surd$ \\
$g_\infty(\vac x,0,0)\sim|\vac x|^{-(d-2)}$ & $\surd$ & $\surd$ &  $\surd$ & ? & $\surd$ \\
\hline
$\xi_L(t=0,0)\sim L^{d/4}$~\cite{Brezin82} & $L$ & $L$ & $ L$ & $L$ & $\surd$\\
$\chi_L(t=0,0)\sim L^{d/2}$~\cite{Binder85} & $L^2$ & $L^2$ & $L^2$ & $\surd$ & $\surd$\\
$  t_L(t=0,0)\sim L^{-d/2}$~\cite{Binder85} & $L^{-1/2}$ & $L^{-1/2}$ & $L^{-1/2}$ & $\surd$ & $\surd$\\
$m_L(t=0,0)\sim L^{-d/4}$ & $L^{-1}$ & $L^{1-d/2}$ & $L^{-1}$ & $\surd$ & $\surd$\\
$g_L(L/2,t=0,0)\sim L^{-d/2}$ & $L^{-(d-2)}$ & $L^{-(d-2)}$ & $L^{-(d-2)}$ & ? & $\surd$ \\
$h_L^{\rm LY}(t=0)\sim L^{-3d/4}$ & $L^{-3}$ & $L^{-(d+2)/2}$ & & & $\surd$ \\
$  t_L^{\rm F}(t=0)\sim L^{-d/2}$ & $L^{-2}$ & $L^{-2}$ & & & $\surd$ \\
\hline
\end{tabular}
\caption{Summary of the evolution of the scaling picture above the upper critical dimension for the $\phi^4$ model.
The first column presents the correct results (FSS predictions are for a system with periodic boundary conditions). 
In the other columns, we give the (incorrect) results predicted when they are different. 
A question mark means that the quantity hasn't been considered in the corresponding scenario.
 $^1$FSS with Landau exponents, $^2$Predictions from  the RG eigenvalues at the Gaussian Fixed Point, $^3$Corrections made by the scenario of Fisher,  $^4$Most of the results presented in this column correspond to BNPY's version of the scenario of Fisher and are checked in Ref.~\cite{Binder85}, $^5$Q scaling. 
 of the results presented in the last column are checked in Ref.~\cite{BERCHE2012115}.
 }
 \label{tab6}
\end{center}
\end{table}

\subsection{Corrections to scaling and crossover}

Equations (\ref{eq-140})--(\ref{eq-202})  are the leading contributions, 
but corrections to scaling can play important roles as well, and somewhat change the picture
(and don't forget the orthogonal G- modes which are always lurking in the background).
The explicit inclusion of corrections to scaling has been introduced analytically  and masterfully checked numerically by Luijten~\cite{PhHLuijten}. 
The role of these corrections could a priori be the origin of a discrepancy that we will discover in Section~\ref{sec9} for systems with {free boundary conditions} (FBC). 
This option will not turn out to be the right one, as we shall see, but we must logically exploit it and, in any case, it is important to explore the corrections in the vicinity of the dominant behaviour.
Rather than assuming different scaling hypotheses for {different} sets of boundary conditions, universality suggests a single {FSS}  behaviour in the thermodynamic limit {that} could leave enough room for crossover effects to take place, and this more severely to manifest in systems with  FBC's. 
Anticipating part of the next section, 
the suspicion in favor of this scenario comes from the numerical results for FBC's which depend very strongly on the manner of carrying out the calculations --- keeping a core of spins only, or removing various boundary submanifolds like corners, edges, surfaces, etc. The resulting exponents scatter around values for which it is hard to find a consistent interpretation. Let us quote a recent work by Lundow~\cite{LUNDOWarxiv}
\begin{quotation}
A system with periodic boundary conditions then has $\chi\propto L^{5/2}$. Deleting the $5L^4$ boundary edges we receive a system with free boundary conditions and now $\chi\propto L^{2}$. In the present work we find that deleting the $L^4$ boundary edges along just one direction is enough to have the scaling $\chi\propto L^{2}$. 
It also appears that deleting $L^3$ boundary edges results in an intermediate scaling, here estimated to $\chi\propto L^{2.275}$.
\end{quotation}

At the risk of repeating ourselves, we will see that this is not the {full} scenario at work in reality but we cannot rule it out  a priori and this is the main reason for this section. 
We also list tables of the leading and first correction exponents for various quantities and various universality classes and this may be helpful for future studies.
So, we will exploit further the formalism introduced previously in Section~\ref{sec6.4}. 

Let us specify the discussion by  {presenting} the case of the susceptibility.
Extracting  the finite-size susceptibility from equation (\ref{eq-134}) in zero magnetic field leads to
\be
\chi_L(  t,0,u)=L^{2\qqq}{\mathscr X}(X) 
\ee
 where the variable $X$ is defined in (\ref{eq-137}). We  call this variable $X_L(  t,u)$ for $b=L$, and its explicit form in terms of $d_{\rm uc}$ and $\qq$ is 
\be
X_L(  t,u)
=L^{2\qqq}  t u^{-\frac {d_{\rm uc}-2}{d_{\rm uc}}}-\tilde p u^{\frac 2{d_{\rm uc}}}
L^{2\qqq+y_u-y_t}
\ee
with ${y_u-y_t} =(4-2d_{\rm uc}\qq)/(d_{\rm uc}-2)<0$.
The first term grows faster with the system size and dominates in the thermodynamic limit for the values of $d>d_{\rm uc}$ considered here.

At the pseudo-critical point, as defined in (\ref{eq-133}), the scaling variable takes the value $X_L(  t_L)=X_0$ and the susceptibility is simply
\be
\chi_L(  t_L,0,u)\simeq {\mathscr X}(X_0)L^{2\qqq},
\ee
with the correct FSS exponent.

Let us suppress the $u$ dependence and consider a simplified expression  $X_L(  t)$ for our forthcoming discussion:
\be
X_L(  t)=A L^{2\qqq}  t -B
L^{4(1-\qqq)/(d_{\rm uc}-2)}.
\ee
{At} the critical point $  t=0$, $X_L(t=0)=-B
L^{y_{\rm corr}}$ and
\be
\chi_L(0,0,u)\simeq {\mathscr X}(-B
L^{y_{\rm corr}})L^{2\qqq}
\ee
with $y_{\rm corr}=y_u-y_t+y_t^*=4(1-\qqq)/(d_{\rm uc}-2)<0$.
The scaling function is regular with an argument $X_L(t=0)\to0$ when $L\to\infty$ which allows to expand ${\mathscr X}(-B
L^{y_{\rm corr}})$ to first order in the vicinity of $X_L(t=0)=0$,
\be
\chi_L(0,0,u)\simeq {\mathscr X}(t=0)L^{2\qqq}-B{\mathscr X}'(t=0)L^{2\qqq+y_{\rm corr}}
{+ \dots} ,\label{eq_206}
\ee
where now $2\qq+y_{\rm corr}=(2\qqq(d_{\rm uc}-4)+4)/(d_{\rm uc}-2)$.
It is very clear that here a crossover takes place, depending on the relative magnitudes of both terms. If the scaling function is such that for small enough sizes the condition
\be
(L/\ell_0)^{4(1-\qqq)/(d_{\rm uc}-2)}\equiv | B
 {\mathscr X}'(t=0)/{\mathscr X}(t=0) | L^{4(1-\qqq)/(d_{\rm uc}-2)}\gg 1\label{eq-165}\ee holds (remember that $4(1-\qqq)/(d_{\rm uc}-2)<0$),
the leading behaviour is governed in this regime by the second term. Instead of $2\qqq$, the FSS exponent measured there becomes closer to the corrected value
$(2\qqq(d_{\rm uc}-4)+4)/(d_{\rm uc}-2)$.
{In the case where $d_{\rm uc}=4$, such as in the Ising model, this leads to $\chi_L(t=0) \sim L^2$ --- precisely the same as what would arise from Landau FSS.}
This cannot hold in the thermodynamic limit, so that this is only an effective exponent, and $2\qqq$ remains the only true FSS exponent for the susceptibility {there}.

The leading corrections to scaling have been studied in the thesis of Flores-Sola~\cite{flore2016},
but equation (\ref{eq_206}) is only the beginning of the expansion, including all sorts of corrections to scaling~\cite{PhysRevB.5.4529}, derived and checked by Luijten~\cite{PhHLuijten}, e.g. 
\bea
\chi_L(t,u)=L^{2y_h^*-d}(a_0&+&a_1L^{y_t^*}[t(1+s_1L^{y_u})+p_1L^{y_u-y_t}]\nnb\\
&+&a_2L^{2y_t^*}[t(1+s_1L^{y_u})+p_1L^{y_u-y_t}]^2+\dots\nnb\\
&+&b_1L^{y_u}\nnb\\&+&b_2L^{2y_u}+\dots\quad).
\eea

\def\tvi{\vrule width 0cm height 4mm depth 2mm}
\begin{table}[ht]
{\footnotesize
\begin{center}
\begin{tabular}{ l l |cc |cc |cc}
\hline
& & \multicolumn{2}{c}{$m_L(0,0,u)$} &   \multicolumn{2}{c}{$\chi_L(0,0,u)$}  &  \multicolumn{2}{c}{$c_L(0,0,u)$}  \\
Model & $\phi^n$  & $y_h^*-d$ & $y_h^*-d+y_{\rm corr}$ & $2y_h^*-d$ & $2y_h^*-d+y_{\rm corr}$ & $2y_  t^*-d$ & $2y_  t^*-d+y_{\rm corr}$\tvi\\
\hline
Magnets, SAW & $\phi^4$  & $\mathcoppa-\frac d2$ &$2-\frac{d}2-\mathcoppa$ & $2\qq$ & {{$2$}} & $4\qq-d$ & $2\qq-d+2$\tvi    \\ 
Percolation & $\phi^3$  & $\mathcoppa-\frac d2$ & {{$1-\frac{d}2$}} & $2\qq$ & $1+\qq$ & $4\qq-d$ & $3\qq-d+1$\tvi   \\
Tricriticality & $\phi^6$ & $\mathcoppa-\frac d2$ &$4-\frac{d}2-3\mathcoppa$ & $2\qq$ & $4-2\qq$ & $4\qq-d$ & {{$4-d$}}\tvi  \\
\hline 
\end{tabular}
\caption{Summary of the crossover expected {for FSS} at the asymptotic critical point for the magnetization, the susceptibility and the specific heat for the $\phi^n$ models. In this table, $y_{\rm corr}=y_u-y_t+y_t^*=4(1-\qq)/(d_{\rm uc}-2)$.}\label{tab9}
\end{center}
}
\end{table}
Similar crossovers are obtained  for the other thermodynamic quantities. In Table~\ref{tab9} we collect the exponents of the leading and first correction term for the magnetization, the susceptibility and the specific heat for the different universality classes that we usually consider in this paper.

\subsection{Corrections in the correlation sector}

The case of the correlations can be discussed in a similar manner.
The assumption (\ref{eq-135}), rewritten here in complete form in zero magnetic field reads as
\be
g_L(\vac x,  t,0,u)=b^{-2x_\phi^*}{\mathscr G}^\pm(b^{-1}\vac x,
b^{y_  t^*}  t u^{-\frac 2n}-\tilde p u^{1-\frac 2n}
b^{y_u-y_  t+y_  t^*},bL^{-1}).\label{eq-199}
\ee
Using the same arguments as above, one can fix e.g. $b=|\vac x|$ to get a first-order expansion at the pseudo-critical point $  t_L$ (unit vector $\vac u=\vac x/|\vac x|$ omitted)
\bea
g_L(\vac x,  t_L,0,u)&=&|\vac x|^{-2x_\phi^*}{\mathscr G}^\pm(X_0,|\vac x|L^{-1})\nnb\\
&\simeq&|\vac x|^{-2x_\phi^*}[
{\mathscr G}^\pm(X_0,0)+|\vac x|L^{-1}{{\mathscr G}^\pm}^{(0,0,1)}(X_0,0)
]\, ,\label{E200}
\eea
and at the asymptotic critical point,
 \bea
g_L(\vac x,0,0,u)&=&|\vac x|^{-2x_\phi^*}{\mathscr G}^\pm(-B|\vac x|^{y_{\rm corr}},|\vac x|L^{-1})\nnb\\
&\simeq&|\vac x|^{-2x_\phi^*}[
{\mathscr G}^\pm(0,0)
-B|\vac x|^{y_{\rm corr}}{{\mathscr G}^\pm}^{(0,1,0)}(0,0)
+|\vac x|L^{-1}{{\mathscr G}^\pm}^{(0,0,1)}(0,0)
].
\nnb
\\
\eea
Here,  ${\mathscr G}^{(0,1,0)}$ for example, means that we take the first derivative of the function ${\mathscr G}$ wrt its second argument.

It is probably easier to study numerically a fixed ratio of the distance $|\vac x|$ to the size of the system, e.g. $\frac 12$. Then one chooses $b=L$ and it comes out
\bea
g_L(L/2,  t_L,0,u)&=&L^{-2x_\phi^*}{\mathscr G}^\pm(\half,X_0,0)\label{E202}\, ,
\\
g_L(L/2,0,0,u)&=&L^{-2x_\phi^*}{\mathscr G}^\pm(\half,-BL^{y_{\rm corr}},0)\nnb\\
&\simeq&L^{-2x_\phi^*}[
{\mathscr G}^\pm(\half,0,0)
-BL^{y_{\rm corr}}{{\mathscr G}^\pm}^{(0,1,0)}(\half,0,0)
].
\label{E203bis}
\eea
While (\ref{E202}) exhibits a pure decay with an exponent $-2x_\phi^*$, (\ref{E203bis}) should display a crossover between an exponent $-2x_\phi^*+y_{\rm corr}$ at small sizes to 
$-2x_\phi^*$ in the thermodynamic limit. We report the corresponding values in Table~\ref{tab10} for the three usual universality classes.

\def\tvi{\vrule width 0cm height 4mm depth 2mm}
\begin{table}[ht]
{\footnotesize
\begin{center}
\begin{tabular}{ l l |cc |cc }
\hline
& & \multicolumn{4}{c}{$g_L(L/2,0,0,u)$}   \\
Model & $\phi^n$  & $2x_\phi^*$ &$\eta^*$ & $2x_\phi^*-y_{\rm corr}$ &$\eta_{\rm corr}$ \tvi\\
\hline
Magnets, SAW & $\phi^4$  & $d-2\qq$  & $2-2\qq$  &$d-2$ & $0$ \tvi    \\ 
Percolation & $\phi^3$ & $d-2\qq$ & $2-2\qq$ &$d-1-\qq$ & $1-\qq $\tvi   \\
Tricriticality & $\phi^6$ & $d-2\qq$ & $2-2\qq$ &$d-4+2\qq$ & $ 2\qq-2$ \tvi  \\
\hline 
\end{tabular}
\caption{Summary of the crossover expected at the asymptotic critical point for the correlation function for the $\phi^n$ models. The corresponding values of $\eta$ exponents, defined by $2x_\phi^*=d-2+\eta^*$ and $2x_\phi^*-y_{\rm corr}=d-2+\eta_{\rm corr}$ are also given.}\label{tab10}
\end{center}
}
\end{table}

As far as we know, the correlation function crossover in terms of distance has not been  {{numerically}} checked yet.

\section{The case of Free Boundary Conditions}\label{sec9}

\subsection{The problem}
We will now devote a moment to the difficult case of free boundary conditions.

To discuss first the easy part, the study of FSS properties of all physical quantities evaluated {\em at the pseudo-critical point} for FBC's agrees with all what was said previously for systems with PBC's. 
We believe that there is now a consensus on this, but what is happening right at the critical point is subject to a lot of discussion and consensus has not yet been obtained. In particular, the energy sector is actually misunderstood as we will see.

In Ref.\cite{BERCHE2012115}, $5$D systems of moderate sizes (up to $L\simeq 30$) were studied at $T_c$.
Besides the full lattice, to mitigate the role of boundary submanifolds which effectively have lower dimensionalities, outer half of the sites were removed in each direction, leaving a core which is genuinely five-dimensional.
The results for the FSS of the susceptibility and the magnetization were not completely conclusive,
\bea
\hbox{core sites}&& \chi_{\rm core}(T_c)\sim L^{1.92},\label{E182}\\
&&m_{\rm core}(T_c)\sim L^{-1.575},\label{E183}\\
\hbox{all sites}&& \chi_{\rm all}(T_c)\sim L^{1.71},\label{E202bis}\\
&&m_{\rm all}(T_c)\sim L^{-1.70}.\label{E203}
\eea
As we said in Ref.\cite{BERCHE2012115},
\begin{quotation} {\em
again, at $T_c$, the data follow neither the Gaussian nor the Q-behaviour}
\end{quotation}
 (Q-behaviour referring there to Q scaling). We had then proposed  that the asymptotic regime was not reached and that the $5$D behaviour was contaminated by the $4$D one of the free surfaces and even by edges, corners, etc. of still lower dimensionalities.
  The perspective offered by the previous section is a tempting solution of the problem. Indeed, the exponent in Eq.~(\ref{E203}), for example, is close to  the value $2-\frac d2-\qq=-1.75$ for the correction exponent of the magnetization in Table~\ref{tab9}. On the other hand, the corresponding exponent in Eq.~(\ref{E202bis}) does not appear to be compatible with the value 2 reported for the correction in the same table. For this very quantity, the core sites susceptibility (\ref{E182}) displays a closer exponent, and that of the magnetization (\ref{E183}) is now right between $\qq-\frac d2=-1.25$ and $2-\frac d2-\qq=-1.75$, the leading and correction exponents (see Table~\ref{tab10}), however, it seems difficult to reach any reliable conclusion on the basis solely of these results.

Lundow and Markstr\"om are experts in simulations and they have studied  this problem intensively, reaching far larger systems (up to $L=160$ in \cite{LUNDOW2014249}) in which the fraction of ``non-bulk'' sites is much smaller. They obtained very accurately leading behaviours and corrections to scaling:
\bea
&&\chi_L(T_c)=0.817\ \!L^2+0.083\ \!L,\label{E177}\\
&&m_L(T_c)=0.230\ \!L^{-3/2}+1.101\ \!L^{-5/2}-1.63\ \!L^{-7/2}.\label{E187}
\eea
On the basis of these results, Lundow and Markstr\"om concluded in favor of a {\em standard FSS behaviour} of the susceptibility at the critical temperature  $\chi_L(T_c)\sim L^2$. The same conclusion, albeit with much lower accuracy, was reached in  \cite{BERCHE2012115} on the basis of (\ref{E182}). There, the use of the word {\em Gaussian} was misleading, since we referred in fact to FSS with Landau exponents, $\chi_L\sim L^{\gamma_{\rm\scriptscriptstyle MFT}/\nu_{\rm\scriptscriptstyle MFT}}$, a conclusion that we will see is {\em not} correct, although $L^2$ is (accidentally) correct.

The case of the magnetization in (\ref{E187}) is a bit more subtle, because the leading exponent $-\frac 32$ {\em is not} the standard Landau FSS (this would be $-\beta_{\rm\scriptscriptstyle MFT}/\nu_{\rm\scriptscriptstyle MFT}=-1$).  
Even more puzzling are the cases of the internal energy and  specific heat at criticality, for which the corrections to scaling were also reported in \cite{LUNDOW2014249}:
\bea
&&e_L(T_c)=0.68-1.01 L^{-1}+0.39 L^{-3/2},\label{206}\\
&&c_L(T_c)=14.69-14.93 L^{-1/3},\label{207}
\eea
but we are still lacking an explanation for these results.

In order to clarify the situation, Wittmann and Young~\cite{PhysRevE.90.062137,WittmannThesis}, and then Flores-Sola et al.~\cite{PhysRevLett.116.115701}
 considered the behaviour of the Fourier modes in a finite system with free boundary conditions.
Following Ref.~\cite{PhysRevB.32.7594}\footnote{An ``underappreciated paper'' as described by Wittmann and Young.}, 
a sine-expansion of the scalar field in the $\phi^4$ action is performed with the boundary conditions 
$\phi({\bf x})=0$ at the free surfaces. 
\be\phi({\bf x})=\sum_{\vac k}\tilde\phi_{\vac k}\psi_{\vac k}
=
\sum_{\bf k}\tilde\phi_{\bf k}\prod_{\alpha=1}^d\sqrt{2/L}\sin k_\alpha x_\alpha,\ee 
where the wave vector components take the values  
$k_\alpha =n_\alpha \pi/(L+1)$, $n_\alpha =1,2,\dots, L$.

 The action takes in ${\bf k}-$space a form  different to that for  PBC's. 
 Distinction should be made between the modes for which all $n_\alpha $-values are odd integers,  which are analogous to the zero mode in the PBC case  (we denote in FBC's their set by ${\cal Q}$),
  and all other modes.
  We discriminate the modes in ${\cal Q}$ because these are the only ones to have the symmetry of the average order parameter  wrt  the center of the lattice  for  FBC systems.  Therefore, these modes may  possibly have a substantial contribution to the equilibrium magnetization.
The remaining modes, for which symmetry reasons exclude any substantial contribution to the average order parameter, are denoted by ${\cal G}$ (like {\em Gaussian}). 
The action now reads as~\cite{PhysRevB.32.7594}
\begin{eqnarray}
  F_{\rm GLW}[\tilde\phi_{\vac k}] 
	 = {\frac 12}
	   \sum_{{\bf k}}
              \Bigl({r_0+c|{\bf k}|^2
						 }\Bigr)\tilde\phi_{\vac k}^2
	  -
		\left({\frac{8}{L}}\right)^{\frac{d}{2}}
		h
		\sum_{\vac k {\in {\cal Q}}}
		\tilde\phi_{\vac k}
		\prod_{\alpha=1}^d{\frac{1}{k_\alpha}}
 \nonumber\\
 +   
{\frac{u}{L^d}}
\sum_{{\bf k}_1,{\bf k}_2,{\bf k}_3,{\bf k}_4}
{
	\Delta_{{\bf k}_1,{\bf k}_2,{\bf k}_3,{\bf k}_4} \tilde\phi_{{{\bf k}_1}}	
\tilde\phi_{{{\bf k}_2}}\tilde	\phi_{{{\bf k}_3}}	\tilde\phi_{{{\bf k}_4}}.
}
	\quad
\label{eq-actionexpandedFBC}
\end{eqnarray}
The quantities $\Delta_i$'s are momentum-conserving factors. Two important differences between Eqs.~(\ref{E162}) and~(\ref{eq-actionexpandedFBC}) are observed. The first one is in the quadratic terms. This  is the source for the different scaling observed between the two types of boundary conditions.
It also appears that the quartic term {in Eq.~(\ref{eq-actionexpandedFBC})} is dangerous  for the modes $\vac k \in {\cal Q}$ only  (due to the restricted summation $\sum_{\vac k {\in {\cal Q}}}$) which couple to $h$.
We henceforth refer to modes for which $u$ is dangerous 
({in particular}, the zero mode  for PBC's and modes with all  $n_\alpha$ odds  for FBC's)
as Q-modes and the remaining ones as  G-modes or Gaussian modes.

With the notation $\tilde m_{\vac k}=\tilde\phi_{\vac k}$, already introduced to represent 
the contribution of a single mode $\bf k$ to the average magnetization, 
one defines the corresponding susceptibility by\footnote{Actually, for the modes which do not contribute to the average magnetization, (i.e. such that $\langle | \tilde m_{\vac k}| \rangle=0$),  the susceptibilities are defined solely by the first terms in  equation (\ref{e_182}).}
\be
\chi_{\bf k}=L^d(\langle \tilde m_{\vac k}^2\rangle-\langle | \tilde m_{\vac k}| \rangle^2).
\label{e_182}
\ee
The equilibrium  magnetization $m$ takes all modes into account and the total susceptibility is defined accordingly. Wittmann and Young confirmed the scaling at $T_c$ for the total susceptibility (for FBC). They found that the single mode susceptibility also obeys a
{\em standard FSS behaviour}, $\chi_{\vac k}\sim L^2$ for the modes
which {\em will not acquire a nonzero magnetization}, namely with the smallest wave-vector with an even $n_\alpha$. We will come back to this in a moment.

An argument in favor of this result, given in~\cite{PhysRevE.90.062137}, follows from (almost) ordinary 
scaling and helps to understand the sense of the word {\em standard } FSS for the authors. 
Actually, Wittmann and Young proposed to use $\bar t=T-T_L$, with $\bar t=t+{\rm const}\ \!\times  L^{-\lambda}$,
 as the temperature-like scaling variable and the starred RG dimensions in the scaling hypothesis for the susceptibility. They wrote then 
$\chi_L(\bar t)=L^{2y_h^*-d}{\mathscr X}(L^{y_t^*}\bar t)$ in zero magnetic field and  the compatibility with 
the bulk behaviour 
$\chi_\infty(t)\sim|t|^{-\gamma_{\rm\scriptscriptstyle MFT}}$ is recovered in the thermodynamic limit by the demand that the asymptotic regime obeys ${\mathscr X}(x)\sim x^{-\gamma_{\rm\scriptscriptstyle MFT}}$ (with $\gamma_{\rm\scriptscriptstyle MFT}=1$ here) for $x\to \infty$, giving $\chi_L(\bar t)\sim L^{2y_h^*-d+y_t^*}(\bar t)^{-1}$.
When $\lambda=d/2$, as it is the case in PBC, this leads at criticality  to $\chi_L(T_c)=L^{d/2}$. On the other hand if $\lambda =2$ (FBC, a point that we comment further below), this amounts to
$\chi_L(T_c)=L^{2}$.

\subsection{Four possible scenarios}

We can however ask why $y_t^*$ and $y_h^*$ have been used in FBC instead of the unstarred exponents, or rather, which ones have been used in fact. Indeed, as we have already noted several times, one cannot disentangle, with the susceptibility, the predictions of {\em standard} (or Landau) FSS from those of the {\em Gaussian Fixed point} FSS. 
The susceptibility is clearly not a good quantity to analyse, and the primary aim of Ref.~\cite{PhysRevLett.116.115701} was to test the more discriminating case of the magnetization.
The  argument of Wittmann and Young for the magnetization would lead to $m_L(T_c)=L^{-\lambda/2}$, hence
$m_L(T_c)=L^{-d/4}$ in PBC and  $m_L(T_c)=L^{-1}$ in FBC.
The argument  is thus equivalent to {\em Landau scaling}, because for FBC, $\lambda=1/\nu_{\rm\scriptscriptstyle MFT}$.

So we are led to the point where one essentially faces four distinct hypotheses among which one has to discriminate (see Table~\ref{table11}).

\begin{table}[th]
\begin{center}
\begin{tabular}{ lll}
\hline
Hypothesis & Scaling of $\chi_L(T_c)$ & Scaling of $m_L(T_c)$ \\
\hline
1. Landau (or {\em standard}) scaling & $L^{\frac{\gamma_{\rm\scriptscriptstyle MFT}}{\nu_{\rm\scriptscriptstyle MFT}}}$   & $L^{-\frac{\beta_{\rm\scriptscriptstyle MFT}}{\nu_{\rm\scriptscriptstyle MFT}}}$ \\
2. Q scaling & $L^{2\qqq}$ & $L^{\qqq-d/2}$ \\
3. G scaling & $L^{2}$ & $L^{1-d/2}$ \\
4. Crossover & $A_\chi L^{2\qqq}-B_\chi L^{2\qqq+\frac{4(1-\qqq)}{d_{\rm uc}-2}}$ & $A_m L^{\qqq-d/2}-B_m L^{\qqq-d/2+\frac{4(1-\qqq)}{d_{\rm uc}-2}}$\\
\phantom{4.\ }\\
\hline
\end{tabular}
\end{center}
\caption{The four scenarios for the FSS of the susceptibility and the magnetization at $T_c$. There, $\qq=d/d_{\rm uc}$, $d_{\rm uc}$ and the $_{\rm\scriptscriptstyle MFT}$ exponents take their respective values for the three different universality classes under consideration. }\label{table11}
\end{table}

In the case of the $\phi^4$ universality class, the options are thus
\begin{description}
\item{---}  prediction 1 (Landau scaling) leads to $\chi_L(T_c)\sim L^2$ and $m_L(T_c)\sim L^{-1}$, 
\item{---} prediction 2 (Q scaling) to $\chi_L(T_c)\sim L^{d/2}$ and $m_L(T_c)\sim L^{-d/4}$, 
\item{---} prediction 3 (G scaling)  to $\chi_L(T_c)\sim L^2$ and $m_L(T_c)\sim L^{-(d-2)/2}$, and 
\item{---} prediction 4 (crossover  to Q scaling) with effective exponents between  $d/2$ and $2$ for $\chi_L$ and between $-d/4$ and $2-3d/4$ for $m$. 
\end{description}
Option 2 is clearly ruled out at $T_c$ by the results of Lundow and Markstr\"om in (\ref{E177}) (as well as by the results of Wittmann and Young). Options 1, 2 and 4 are all compatible with the results measured for the susceptibility. For the 5D model, it is easy to discriminate, with the magnetization,  between option 1 (which predicts the value $-1$), option 3 (prediction $-1.5$) and option 4 (prediction between $-1.25$ and $-1.75$). Equation (\ref{E187}) is clearly in favor of option 3,  but does not rule out option 4.

\subsection{Towards a GFP scaling at $T_c$}
In Ref.~\cite{PhysRevLett.116.115701} the magnetization of the LRIM was investigated to tune the parameter $\sigma$ of the interaction decay. 
In the thesis of Flores-Sola~\cite{flore2016} results are reported for various values of $\sigma $ and  of $d$. 
The expectation from GFP scaling is an exponent $-\beta_\G/\nu_\G$ with $\beta_\G=(d-\sigma)/(2\sigma)$ and $\nu_\G=1/\sigma$, while Q scaling predicts an exponent $-\qq\beta_\MFT/\nu_\MFT=-\qq$.

Other quantities (temperature shift, correlation length, correlation function) are also reported in~\cite{flore2016} 
and, although perfectible,  all numerical results at $T_c$ support option 3 above (and confirm Q scaling at $T_L$).

The picture now is the following: 
{\emph{Either}} we study physical quantities which are related to the {Q modes} or those related to the G modes.
In the first case, which is expected at $T_L$, the DIV has to be taken into account and Q scaling rules (option 2 in Table~\ref{table11}).
In the second case, which holds at $T_c$,  {$u$ is not dangerous} and the physics is controlled by the GFP (option 3 in Table~\ref{table11}).
The exponents there are those  collected in Table~\ref{tabGaussian}, {\it and not Landau exponents} collected in Table~\ref{tab2}.
There still remains to understand why quantities at $T_c$ averaged over all modes in FBC obey G scaling rather than Q scaling. 
This is surprising, since average properties are dominated by Q modes. We believe that the reason is entirely due to the behaviour of the shift of the pseudo-critical temperature $t_L=T_c-T_L$. 
It was indeed shown already in Ref.~\cite{PhysRevB.32.7594} that
 \be
 t_L\sim L^{-2}
 \ee
 for the Ising model with FBC's  above its upper critical dimension, while $t_L\sim L^{-d/2}$ for PBC's. 
 In both cases, the rounding is still governed by the exponent $d/2$, hence we observe that in FBC's, the shift is much larger than the rounding, and $T_c>T_L$. This means that at $T_c$, the finite system is effectively disordered. As a consequence, the average order parameter profile {\em at $T_c$} is essentially vanishing and Q modes, like G modes have negligible contributions at that temperature, like in the high temperature phase. We believe that this could be the origin of the observed FSS for FBC's and, in a sense,  the variable $u$ is not able to render the zero mode order parameter dominant.

This could be the end of the story but for the fact that the above interpretations in the energy sector (internal energy and specific heat) don't match the recent numerical observations of Lundow and Markstr\"om in (\ref{E177}).
 These measurements disagree with all three options: Q scaling, G scaling and also the less likely Landau scaling. 
  The fourth option of a crossover does not do the job {either.}  
 For the internal energy, it would predict a competition between terms in $L^{-2.5}$ and $L^{-3}$, and for the specific heat, $L^0$ would compete with $L^{-0.5}$.
 Of course the numerical determination of the internal energy and  of the specific heat is known to be made difficult by the presence of important regular contributions, so there is still work to do there.

A last point to emphasize concerns the case of percolation for which also competing theories have been elaborated. Let us remind the reader that the analog of the susceptibility for percolation is the average size of finite clusters $S(p)$ (with $p$ the probability that a site or a bond is occupied) and that a central quantity which plays the role of the free energy density is the density of finite clusters $K(s,p)$. The phase transition which stands out the two phases (existence vs non existence of spanning cluster(s)) occurs at a probability $p_\infty$ and the analog of the reduced temperature there is $\epsilon= |p-p_\infty|$. The homogeneous form of the free energy density then takes the standard form $K(s,p)=b^{-d}{\mathscr F}(\kappa b^{D_f},\epsilon b^{1/\nu})$. Here, the analog of the magnetic field is $\kappa=s^{-1}-(s^{\rm max})^{-1}$ with $s^{\rm max}$ the typical mass of the largest clusters which scales near $p_\infty$ like $s^{\rm max}_\infty(p)\sim\epsilon^{-\frac 1\sigma}\sim(\xi_\infty)^{D_f}$ with a fractal dimension $D_f$, hence $\sigma=1/(\nu D_f)$. $D_f$ is the analog of $y_h$ in our discussion. 

Above $d_{\rm uc}=6$, Antonio Coniglio proposed a scenario according to which there is proliferation of interpenetrating spanning clusters \cite{Coniglio}, but $D_f$ is stuck to its value $D_{\rm uc}=4$ at the upper critical dimension. The resulting free energy density for finite systems takes the form
\be
K_L(s,p)=b^{-(d-X)}\mathscr F(b^{D_{\rm uc}}\kappa,b^{1/\nu_c}\epsilon, bL^{-1})
\ee
with $X=d-d_{\rm uc}$ the exponent which measures the proliferation of spanning clusters. According to the literature (see e.g. \cite{Kenna_2017}), this situation is encountered in FBC's. It corresponds exactly to scenario 1 in Table \ref{table11}, i.e. Landau FSS, a case which was excluded for the IM universality class! Finite systems with PBC's on the other hand seem to obey a different ansatz, with no proliferation of percolating (wrapping) clusters of fractal dimension $D_f=D^*=2d/3$ \cite{Heydenreich}. This agrees with scenario 2 of Q FSS in Table \ref{table11}. 

Clearly, progress has been made, but the case of FBC's is not yet fully understood and  further studies are underway \cite{submitted}.

\section{Conclusions}
Because they are now endemic in studies of phase transitions and critical phenomena,  
it is almost forgotten that the notation for the six main critical exponents at one point required standardisation. 
It was Michael Fisher who assigned the labels
$\alpha$, $\beta$, $\gamma$, $\delta$, $\eta$ and $\nu$ to the observables listed in Eqs.(\ref{eq-1}--\ref{eq-5}) in the 1960's~\cite{standard}.
These are linked by the four scaling relations (\ref{oldhyper})-(\ref{oldeta}), 
the last of which   was often considered to fail above the upper critical dimension.
The alluring simplicity of the (correct) mean-field description of scaling there 
hid subtleties that appeared to undermine the renormalization group itself, accurately described in Ref.\cite{Papathanakos}, for example, as  ``plagued by a number of persistent problems''.

In the 1980s, Michael Fisher took the first and most important steps to rescue the situation when he identified ``the only flaw in the original argument was a failure to recognize and allow for possible singular behaviour of the scaling function.''
In the spirit of Occam's razor, and fixing only that which appeared to be broken,   
he addressed the free energy sector in the thermodynamic limit, identifying some irrelevant variables as dangerous. 
While ``the renormalization group framework ha[d] been preserved intact'', 
hyperscaling was sacrificed, its lack of ubiquity only a relatively mild discomfort.

Finite-size scaling had more enigmatic undertones in high dimensions, however.
To quote some of the leading protagonists at the turn of the century  \cite{Binderquote}
``although 
\dots 
all exponents are known, 
\dots
and in principle very complete analytical calculations are possible, the
existing theories clearly are not so good.''
A number of ingenious and even elegant formalisms  had been developed, reflecting 
``the progress and setbacks inherent to the evolution of science'' alluded to above.
Notable amongst these were BNPY's extension of Fisher's dangerous concepts to finite systems \cite{PhysRevB.31.1498}; Binder's thermodynamic length \cite{Binder85}; Coniglio's
proliferating  spanning clusters in percolation \cite{Coniglio} and Luijten's and  Bl\"ote's ~\cite{PhHLuijten} inclusions of corrections to scaling.
The extension of these concepts to the correlation sector led to the introduction of a new exponent  $\mathcoppa$~\cite{CMP2013} which, like the story outlined above, has an enigmatic character; 
although it is a finite-size concept it is needed to recover hyperscaling for the thermodynamic limit 
as Eq.(\ref{hyperhyper}). 

However, the puzzles of FSS above the upper critical dimension are  still not fully resolved and there is plenty of room for further explorations.
We have essentially two different types of behaviour which are partially controlled by the value of the order parameter. 
The two behaviours are governed either by the pure Gaussian Fixed Point (G scaling), or by the same fixed point, but contaminated by the Dangerous Irrelevant Variable (Q scaling).
For physical quantities which do not depend on this variable, all exponents are ordinary Gaussian exponents, and FSS is like $Q_L\sim L^{-\rho_{\rm\scriptscriptstyle G}/\nu_{\rm\scriptscriptstyle G}}$. For other quantities, affected by the DIV, scaling is either like 
$Q_L\sim L^{-\rho_{\rm\scriptscriptstyle G}/\nu_{\rm\scriptscriptstyle G}}$, or like $Q_L\sim L^{-\qqq\rho_{\rm\scriptscriptstyle MFT}/\nu_{\rm\scriptscriptstyle MFT}}$. The second case is more generic, but when 
 the order parameter (a priori ``contaminated'') is strongly bounded to zero by the boundary conditions, there is no room for the DIV to develop and the first case is observed. This is what happens specifically at $T_c$ in systems with FBC's.
 However, this discussion  should be moderated by the case of percolation which does not yet fully fit in this picture \cite{submitted}.
 
 As stated, this is not a review. Nor is it a prediction of the future.
 It is the story of what we consider as being the most {important} developments in the adaption of RG for high dimensions. 
 Still, having exposed the seemingly inert mean-field realm of high dimensions as a rich ground for fundamental research, we  can anticipate new, important and fascinating developments in the future. E.g., while we were finishing these lecture notes, one such paper appeared concerning the extension of the Q FSS formalism to quantum phase transitions~\cite{QuantumQFSS}. 
 Suffice to say that hyperscaling can no longer be described as ``failing'' above the  upper critical dimension. Instead it is rescued by the emergence of $\mathcoppa$ which can take its place amongst the pantheon of critical exponents that were christened by the same authoritative figure to which this paper is dedicated.

\section*{Acknowledgements}
We would like to thank  Emilio Flores-Sola, {former  student of B.B. and R.K.,} who  made his PhD in co-advisory between Coventry University and the Universit\'e de Lorraine.

Yu.H. acknowledges support of the JESH mobility program of the Austrian Academy
of Sciences and hospitality of the Complexity Science Hub Vienna when finalizing this
paper.

As we finished preparing this manuscript, 
we learned of the passing of Michael Fisher through the greatest phase transition of them all.
We take the opportunity to express our thanks for all he has done to enlighten all of us from the very start to the very end of his long and fruitful career.

We take this opportunity to express our indignation at the Russian invasion of Ukraine and the distorted view of history that preceded it.
We refer the reader to Ref.\cite{bylyny} for an analysis of medieval Ukrainian literature which refutes that view and concludes:
 “Thus the Kyiv cycle of east Slavic epic narratives falls nicely within the European tradition in network terms — it is in many ways like Ireland’s heroic tradition and Iceland’s social ones.”
 Finally, express our solidarity with the Ukrainian people and people the world over who actively strive to uphold the most noble principles of  academia and of humanity.

\bibliography{references}

\begin{thebibliography}{10}
\providecommand{\url}[1]{\texttt{#1}}
\providecommand{\urlprefix}{URL }
\expandafter\ifx\csname urlstyle\endcsname\relax
  \providecommand{\doi}[1]{doi:\discretionary{}{}{}#1}\else
  \providecommand{\doi}{doi:\discretionary{}{}{}\begingroup
  \urlstyle{rm}\Url}\fi
\providecommand{\eprint}[2][]{\url{#2}}

\bibitem{Privman1983}
V.~Privman and M.~E. Fisher,
\newblock \emph{Finite-size scaling of the correlation length above the upper
  critical dimension in the five-dimensional ising model},
\newblock J. Stat. Phys. \textbf{33}, 385 (1983),
\newblock \doi{10.1007/BF01009803}.

\bibitem{FisherStellenbosch}
M.~Fisher,
\newblock \emph{Scaling, Universality and the Renormalization Group theory},
  chap.~1, pp. 1--137,
\newblock Springer-Verlag Berlin Heidelberg,
\newblock \doi{110.1007/3-540-12675-9},
\newblock Proceedings of the Summer School Held at the University of
  Stellenbosch, South Africa January 18–29, 1982 (1983),
  \eprint{https://www.springer.com/gp/book/9783540126751}.

\bibitem{doi:10.1142/9789814632683_0001}
R.~Kenna and B.~Berche,
\newblock \emph{Scaling and Finite-Size Scaling above the Upper Critical
  Dimension}, chap.~1, pp. 1--54,
\newblock World Scientific, Singapore,
\newblock \doi{10.1142/9789814632683_0001} (2015),
  \eprint{https://www.worldscientific.com/doi/pdf/10.1142/9789814632683_0001}.

\bibitem{FisherHistorical}
M.~Fisher,
\newblock \emph{Notes, definitions, and formulas for critical point
  singularities},
\newblock In Critical phenomena. Proceedings of a conference held in
  Washington, DC, April 1965, ed. M.S. Green and J.V. Sengers. National Bureau
  of Standards (1966).

\bibitem{BERCHE20107}
B.~Berche, P.~Butera and L.~N. Shchur,
\newblock \emph{Logarithmic corrections and universal amplitude ratios in the
  4-state potts model},
\newblock Physics Procedia \textbf{7}, 7 (2010),
\newblock \doi{https://doi.org/10.1016/j.phpro.2010.09.039},
\newblock Computer Simulation Studies in Condensed Matter Physics XX, CSP-2007.

\bibitem{PrivmanEtAl}
V.~Privman, P.~Hohenberg and A.~Aharony,
\newblock \emph{Universal critical-point amplitude relations},
\newblock In Phase Transitions and Critical Phenomena, vol. 14, ed. by C. Domb,
  J.L. Lebowitz. Academic, New York (1991).

\bibitem{Kennalog}
R.~Kenna,
\newblock \emph{Universal scaling relations for logarithmic-correction
  exponents}, chap.~1, pp. 1--46,
\newblock World Scientific, Singapore,
\newblock \doi{https://doi.org/10.1142/9789814417891_0001} (2012),
  \eprint{https://www.worldscientific.com/doi/10.1142/9789814417891_0001}.

\bibitem{KennaPRL1}
R.~Kenna, D.~Johnston and W.~Janke,
\newblock \emph{Scaling relations for logarithmic corrections},
\newblock Phys. Rev. Lett. \textbf{96}, 115701 (2006).

\bibitem{KennaPRL2}
R.~Kenna, D.~Johnston and W.~Janke,
\newblock \emph{Self-consistent scaling theory for logarithmic-correction
  exponents},
\newblock Phys. Rev. Lett. \textbf{97}, 155702 (2006).

\bibitem{Wi65a}
B.~Widom,
\newblock \emph{Surface tension and molecular correlations near the critical
  point},
\newblock J. Chem. Phys. \textbf{43}, 3892 (1965).

\bibitem{Wi65b}
B.~Widom,
\newblock \emph{Equation of state in the neighborhood of the critical point},
\newblock J. Chem. Phys. \textbf{43}, 3898 (1965).

\bibitem{Gr67}
R.~Griffiths,
\newblock \emph{Thermodynamic functions for fluids and ferromagnets near the
  critical point},
\newblock Phys. Rev. \textbf{158}, 176 (1967).

\bibitem{Ka66}
L.~Kadanoff,
\newblock \emph{Scaling laws for ising models near $t_c$},
\newblock Physics \textbf{2}, 263 (1966).

\bibitem{EsFi63}
J.~Essam and M.~Fisher,
\newblock \emph{Pad{\'{e}} approximant studies of the lattice gas and ising
  ferromagnet below the critical point},
\newblock J. Chem. Phys. \textbf{38}, 802 (1963).

\bibitem{Wi64}
G.~Rushbrooke,
\newblock \emph{Degree of the critical isotherm},
\newblock J. Chem. Phys. \textbf{41}, 1633 (1964).

\bibitem{Fi64sc}
M.~Fisher,
\newblock \emph{Correlation functions and the critical region of simple
  fluids},
\newblock J. Math. Phys. \textbf{5}, 944 (1964).

\bibitem{Jo67}
B.~Josephson,
\newblock \emph{Inequality for the specific heat: I. derivation},
\newblock Proc. Phys. Soc. \textbf{92}, 269 (1967).

\bibitem{Ru63}
G.~Rushbrooke,
\newblock \emph{On the thermodynamics of the critical region for the ising
  problem},
\newblock J. Chem. Phys. \textbf{39}, 842 (1963).

\bibitem{Gr65}
R.~Griffiths,
\newblock \emph{Thermodynamic inequality near the critical point for
  ferromagnets and fluids},
\newblock Phys. Rev. Lett. \textbf{14}, 623 (623-624).

\bibitem{BuGu69}
M.~Buckingham and J.~Gunton,
\newblock \emph{Correlations at the critical point of the ising model},
\newblock Phys. Rev. \textbf{178}, 848 (1969).

\bibitem{Fi69inequalities}
M.~Fisher,
\newblock \emph{Rigorous inequalities for critical-point correlation
  exponents},
\newblock Phys. Rev. \textbf{180}, 594 (1969).

\bibitem{PatashinskiPokrovski}
A.~Z. Patashinskii and V.~L. Pokrovsky,
\newblock \emph{Behavior of ordered systems near the transition point},
\newblock Sov. Phys. JETP \textbf{23}, 292 (1966).

\bibitem{PhysicsPhysiqueFizika.2.263}
L.~P. Kadanoff,
\newblock \emph{Scaling laws for ising models near ${T}_{c}$},
\newblock Physics Physique Fizika \textbf{2}, 263 (1966),
\newblock \doi{10.1103/PhysicsPhysiqueFizika.2.263}.

\bibitem{Berche_2013}
B.~Berche, P.~Butera and L.~N. Shchur,
\newblock \emph{The two-dimensional 4-state potts model in a magnetic field},
\newblock Journal of Physics A: Mathematical and Theoretical \textbf{46}(9),
  095001 (2013),
\newblock \doi{10.1088/1751-8113/46/9/095001}.

\bibitem{cha95}
P.~M. Chaikin and T.~C. Lubensky,
\newblock \emph{Principles of Condensed Matter Physics},
\newblock Cambridge University Press, Cambridge (1995).

\bibitem{Gunton1973RenormalizationGI}
M.~E. Fisher,
\newblock \emph{Renormalization group in critical phenomena and quantum field
  theory : proceedings of a conference, {Temple} university, may 29-31, 1973}
  (1974).

\bibitem{Brezin82}
E.~Brézin,
\newblock \emph{An investigation of finite size scaling},
\newblock J. Phys. France \textbf{43}, 15 (1982),
\newblock \doi{10.1051/jphys:0198200430101500}.

\bibitem{Binder85}
K.~Binder,
\newblock \emph{Critical properties and finite-size effects of the
  five-dimensional ising model},
\newblock Z. Physik B - Condensed Matter \textbf{61}, 13 (1985),
\newblock \doi{10.1007/BF01308937}.

\bibitem{BREZIN1985867}
E.~Brézin and J.~Zinn-Justin,
\newblock \emph{Finite size effects in phase transitions},
\newblock Nuclear Physics B \textbf{257}, 867 (1985),
\newblock \doi{https://doi.org/10.1016/0550-3213(85)90379-7}.

\bibitem{Rickwardt}
C.~Rickwardt, P.~Nielaba and K.~Binder,
\newblock \emph{A finite size scaling study of the five-dimensional ising
  model},
\newblock Annalen der Physik \textbf{506}(6), 483 (1994),
\newblock \doi{https://doi.org/10.1002/andp.19945060606}.

\bibitem{PhysRevB.71.174438}
J.~L. Jones and A.~P. Young,
\newblock \emph{Finite-size scaling of the correlation length above the upper
  critical dimension in the five-dimensional ising model},
\newblock Phys. Rev. B \textbf{71}, 174438 (2005),
\newblock \doi{10.1103/PhysRevB.71.174438}.

\bibitem{BERCHE2012115}
B.~Berche, R.~Kenna and J.-C. Walter,
\newblock \emph{Hyperscaling above the upper critical dimension},
\newblock Nuclear Physics B \textbf{865}(1), 115 (2012),
\newblock \doi{https://doi.org/10.1016/j.nuclphysb.2012.07.021}.

\bibitem{PhysRevB.31.1498}
K.~Binder, M.~Nauenberg, V.~Privman and A.~P. Young,
\newblock \emph{Finite-size tests of hyperscaling},
\newblock Phys. Rev. B \textbf{31}, 1498 (1985),
\newblock \doi{10.1103/PhysRevB.31.1498}.

\bibitem{Caracciolo01}
S.~Caracciolo, A.~Gambassi, M.~Gubinelli and A.~Pelissetto,
\newblock \emph{Finite-size correlation length and violations of finite-size
  scaling},
\newblock Eur. Phys. J. B \textbf{20}, 255 (2001).

\bibitem{PhysRevLett.76.1557}
E.~Luijten and H.~W.~J. Bl\"ote,
\newblock \emph{Finite-size scaling and universality above the upper critical
  dimensionality},
\newblock Phys. Rev. Lett. \textbf{76}, 1557 (1996),
\newblock \doi{10.1103/PhysRevLett.76.1557}.

\bibitem{PhysRevLett.76.3662.3}
E.~Luijten and H.~W.~J. Bl\"ote,
\newblock \emph{Finite-size scaling and universality above the upper critical
  dimensionality},
\newblock Phys. Rev. Lett. \textbf{76}, 3662 (1996),
\newblock \doi{10.1103/PhysRevLett.76.3662.3}.

\bibitem{PhHLuijten}
E.~Luijten,
\newblock \emph{Interaction Range, Universality and the Upper Critical
  Dimension},
\newblock Ph.D. thesis,
\newblock Doctoral thesis, Delft University Press (1997).

\bibitem{flore2016}
E.~J. Flores-Sola,
\newblock \emph{Finite-size scaling above the upper critical dimension},
\newblock Ph.D. thesis,
\newblock Thèse de doctorat dirigée par Berche, Bertrand et Kenna, Ralph
  Physique Université de Lorraine, Coventry University, 2016 (2016).

\bibitem{PhysRev.87.404}
C.~N. Yang and T.~D. Lee,
\newblock \emph{Statistical theory of equations of state and phase transitions.
  i. theory of condensation},
\newblock Phys. Rev. \textbf{87}, 404 (1952),
\newblock \doi{10.1103/PhysRev.87.404}.

\bibitem{PhysRev.87.410}
T.~D. Lee and C.~N. Yang,
\newblock \emph{Statistical theory of equations of state and phase transitions.
  ii. lattice gas and ising model},
\newblock Phys. Rev. \textbf{87}, 410 (1952),
\newblock \doi{10.1103/PhysRev.87.410}.

\bibitem{FisherZeros}
M.~Fisher,
\newblock \emph{The nature of critical points}, vol.~7c, pp. 1--159,
\newblock University of Colorado Press (1965).

\bibitem{Wu}
F.~Wu,
\newblock \emph{Professor c.n. yang and statistical mechanics},
\newblock International Journal of Modern Physics B \textbf{22}, 1899 (2008).

\bibitem{ITZYKSON1983415}
C.~Itzykson, R.~Pearson and J.~Zuber,
\newblock \emph{Distribution of zeros in ising and gauge models},
\newblock Nuclear Physics B \textbf{220}(4), 415 (1983),
\newblock \doi{https://doi.org/10.1016/0550-3213(83)90499-6}.

\bibitem{PhysRevE.90.062137}
M.~Wittmann and A.~P. Young,
\newblock \emph{Finite-size scaling above the upper critical dimension},
\newblock Phys. Rev. E \textbf{90}, 062137 (2014),
\newblock \doi{10.1103/PhysRevE.90.062137}.

\bibitem{PhysRevLett.116.115701}
E.~Flores-Sola, B.~Berche, R.~Kenna and M.~Weigel,
\newblock \emph{Role of fourier modes in finite-size scaling above the upper
  critical dimension},
\newblock Phys. Rev. Lett. \textbf{116}, 115701 (2016),
\newblock \doi{10.1103/PhysRevLett.116.115701}.

\bibitem{PhysRevB.56.8945}
E.~Luijten and H.~W.~J. Bl\"ote,
\newblock \emph{Classical critical behavior of spin models with long-range
  interactions},
\newblock Phys. Rev. B \textbf{56}, 8945 (1997),
\newblock \doi{10.1103/PhysRevB.56.8945}.

\bibitem{CMP2013}
R.~Kenna and B.~Berche,
\newblock \emph{A new critical exponent 'coppa' and its logarithmic counterpart
  'hat coppa'},
\newblock Condensed Matter Physics \textbf{16}, 23601:1 (2013),
\newblock \doi{10.5488/CMP.16.23601}.

\bibitem{Josephson67}
B.~Josephson,
\newblock \emph{Inequality for the specific heat: I. derivation},
\newblock Proc. Phys. Soc. \textbf{92}, 269 (1967).

\bibitem{lotsfq}
Z.~Merdan and D.~Gokbel-Keklikoglu,
\newblock \emph{The test of a new critical exponent by using ising model on the
  creutz cellular automaton},
\newblock Acta Physica Polonica A \textbf{133}, 1200 (2018).

\bibitem{Deger2020}
A.~Deger and C.~Flindt,
\newblock \emph{Lee-yang theory of the curie-weiss model and its rare
  fluctuations},
\newblock Physical Review Research \textbf{2}, 033009 (2020).

\bibitem{LUNDOWarxiv}
P.~Lundow,
\newblock \emph{Boundary effects on finite-size scaling for the 5-dimensional
  ising model},
\newblock arxiv.org/2103.08695  (2021).

\bibitem{PhysRevB.5.4529}
F.~J. Wegner,
\newblock \emph{Corrections to scaling laws},
\newblock Phys. Rev. B \textbf{5}, 4529 (1972),
\newblock \doi{10.1103/PhysRevB.5.4529}.

\bibitem{LUNDOW2014249}
P.~Lundow and K.~Markström,
\newblock \emph{Finite size scaling of the 5d ising model with free boundary
  conditions},
\newblock Nuclear Physics B \textbf{889}, 249 (2014),
\newblock \doi{https://doi.org/10.1016/j.nuclphysb.2014.10.011}.

\bibitem{WittmannThesis}
M.~C. Wittmann,
\newblock \emph{Nature of the spin-glass phase in models with long-range
  interactions},
\newblock Ph.D. thesis,
\newblock Doctoral thesis, UC Santa Cruz (2015).

\bibitem{PhysRevB.32.7594}
J.~Rudnick, G.~Gaspari and V.~Privman,
\newblock \emph{Effect of boundary conditions on the critical behavior of a fi-
  nite high-dimensional ising model},
\newblock Phys. Rev. B \textbf{32}, 7594 (1985),
\newblock \doi{10.1103/PhysRevB.32.7594}.

\bibitem{Coniglio}
A.~Coniglio,
\newblock \emph{Shapes, Surfaces and Interfaces in Percolation Clusters},
\newblock Springer Proceedings in Physics, Vol. 5: Physics of Finely Divided
  Matter, by M.~Daoud, N.~Boccara (Eds.). Springer Verlag,
\newblock Proc. of Les Houches Cong. on Physics of Finely Divided Matter
  (1985).

\bibitem{Kenna_2017}
R.~Kenna and B.~Berche,
\newblock \emph{Universal finite-size scaling for percolation theory in high
  dimensions},
\newblock Journal of Physics A: Mathematical and Theoretical \textbf{50}(23),
  235001 (2017),
\newblock \doi{10.1088/1751-8121/aa6bd5}.

\bibitem{Heydenreich}
M.~Heydenreich and R.~van~der Hofstad,
\newblock \emph{Progress in high-dimensional percolation and random graphs},
\newblock Centre de Recherches Mathématiques, Montreal, QC. Springer (2017).

\bibitem{submitted}
K.~R. Ellis, Tim and B.~Berche,
\newblock \emph{The fifty-year quest for universality in percolation theory in
  high dimensions},
\newblock Submitted  (2021).

\bibitem{standard}
M.~Fisher,
\newblock \emph{Notes, Definitions, and Formulas for Critical Point
  Singularities}, pp. 21--25,
\newblock U.S. Govt. Printing Off. (1966),
  \eprint{https://www.worldscientific.com/doi/pdf/10.1142/9789814632683_0001}.

\bibitem{Papathanakos}
V.~Papathanakos,
\newblock \emph{Finite-size effects in high-dimensional statistical mechanical
  systems: The ising model with periodic boundary conditions},
\newblock PhD thesis, Princeton University  (2006).

\bibitem{Binderquote}
K.~Binder, E.~Luijten, N.~W. M.~M\"{u}ller and H.~Bl\"ote,
\newblock \emph{Monte carlo investigations of phase transitions: status and
  perspectives},
\newblock Physica A \textbf{281}, 112 (2000).

\bibitem{QuantumQFSS}
A.~Langheld, J.~Koziol, P.~Adelhardt, S.~Kapfer and K.~Schmidt,
\newblock \emph{Scaling at quantum phase transitions above the upper critical
  dimension},
\newblock submitted to SciPost Physics  (2022).

\bibitem{bylyny}
P.~Sarkanych, N.~Fedorak, Y.~Holovatch, P.~MacCarron, J.~Yose and R.~Kenna,
\newblock \emph{Network analysis of the kyiv bylyny cycle --- east slavic epic
  narratives},
\newblock https://arxiv.org/abs/2203.10399 and submitted to ACS  (2022).

\end{thebibliography}

\end{document}